\documentclass[dvipdfmx]{jpsj3}
\usepackage{txfonts}
\usepackage{color}

\title{
Finite-size scaling analysis of binary stochastic processes and 
universality classes of information cascade phase transition
}

\author{Shintaro Mori
$^1$\thanks{mori@sci.kitasato-u.ac.jp}
 and Masato Hisakado$^2$}
\inst{$^1$
Department of Physics, Kitasato University \\ 
1-15-1, Kitasato, Sagamihara, Kanagawa 252-0373, Japan \\
$^2$
Financial Services Agency\\
Kasumigaseki 3-2-1, Chiyoda-ku, Tokyo 100-8967, Japan
} %\\

\abst{
We propose a finite-size scaling 
analysis of binary stochastic processes $X(t)\in \{0,1\}$
based on the second moment correlation length $\xi$ for  the 
autocorrelation function $C(t)$.
The purpose is to  clarify the critical properties 
and provide a new data analysis method for information 
cascades. 
As a simple model to represent the different behaviors of 
subjects in information cascade experiments, we assume
 that 
$X(t)$ is a mixture of an independent random variable 
that takes 1 with  probability $q$ and 
 a random variable that depends on the ratio $z$ of the variables 
 taking 1 among recent $r$ variables.
 We consider two types of the probability $f(z)$ that 
 the latter takes 1: (i) analog [$f(z)=z$] and
(ii) digital [$f(z)=\theta(z-1/2)$]. 
We study the universal functions of scaling for $\xi$ and  
the integrated correlation time $\tau$.
For finite $r$, $C(t)$ decays exponentially as a function of $t$, and 
there is only one stable renormalization group (RG) fixed point. 
In the limit $r\to \infty$, where $X(t)$ depends on all the previous 
variables, $C(t)$ in model (i) 
obeys a power law,  and the system 
becomes scale invariant. 
In model (ii) with $q\neq 1/2$, 
there are  two stable RG fixed points, which correspond 
to the ordered and disordered phases of the information 
cascade phase transition with critical exponents 
$\beta=1$ and $\nu_{||}=2$.
}

\begin{document}
\maketitle

\section{\label{sec:introdunction}Introduction}
 Collective phenomena have attracted considerable 
 interest and remain an attractive research subject.
 They are ubiquitous 
 in physical, biological, and social systems 
\cite{Sta:1971,Man:2008,Cas:2009,Gal:2008}. 
Among them, the correlated binary sequence is an important research subject. 
The correlated random walk, quantum walk and P\'{o}lya urn 
process 
are examples of 
the correlated binary sequence 
\cite{Kem:2003,Boh:2000,Pol:1931,Hod:2004,His:2006,Usa:2003,Val:2007}.  
The update sequence of spins in the 
kinetic Ising model is also a correlated binary 
sequence \cite{Gla:1963,Pri:1997}. These systems have rich mathematical 
structures, including 
scale invariance and 
phase transitions \cite{Sta:1971,Sta:1999,New:2005,Hin:2000,Odo:2004}. 

Scale invariance in a binary sequence is usually
defined in terms of the 
scaling behavior of the variance of 
displacements with respect to the length \cite{Kan:1994,Sta:1995}.
If the variance obeys a power law of the length 
and the exponent is greater than 1, it is called superdiffusive 
behavior. Superdiffusion can be attributed to long-range positive 
correlations and 
can be seen in many 
strongly correlated binary sequences in nature 
\cite{Kan:1994}. Coarse-grained DNA strings, 
written texts, and financial data are 
 examples. To explain the scale invariance or 
superdiffusive behavior, a binary stochastic process with a long 
memory has been proposed \cite{Usa:2003,Hod:2004}. 
It is related to a safety campaign problem. 
In the P\'{o}lya--Friedman urn process \cite{Pol:1931,Hui:2008}, 
the probability for a unit bit 
in a binary string depends linearly on the fraction $z$ 
of the unities preceding it.
It shows a dynamical phase transition between  
normal diffusion phase with an exponent of 1 and 
the superdiffusion phase with an exponent greater than 1 
\cite{Hod:2004,Hui:2008}. 
If the memory length is finite and the random variables depend on 
the recent $r$ variables \cite{Usa:2003}, the model is known as 
Kirman's ant model \cite{Kir:1993,Lux:1995}.
It was introduced to describe the switching 
 process of herding behavior in  financial markets. In the model, 
 the number of variables taking 1 among the $r$ variables oscillates 
 randomly \cite{Kir:1993,Kir:2010}. 
 On the basis of the model, the key stylized facts
 in the statistical properties of financial markets have been 
 explained \cite{Alf:2005}.  

The Ising model is a representative example of the phase 
transition of binary variables \cite{Sta:1971}. 
It shows an order-disorder phase 
transition in the thermodynamic limit.
The kinetic Ising model is a single spin update 
stochastic process of the Ising model \cite{Gla:1963}.
As the system approaches the critical point, the relaxation time 
diverges, and the dynamic exponent $Z$ is introduced to classify the 
universality classes of dynamical critical phenomena in 
addition to the basic exponents
\cite{Sta:1971,Odo:2004}.  
The voting model is  a generalized P\'{o}lya urn  
process \cite{Hil:1980,Mor:2015} and describes the 
sequential voting process of subjects (voters) 
in information cascade experiments 
\cite{Bik:1992,Dev:1996} and a race-track betting market 
\cite{Mor:2010}. 
In an information cascade experiment,  
there are two options and 
subjects choose 
options one by one. They can observe the previous subjects' 
choices,
and the observations affect their own choices 
\cite{And:1997,Kub:2004,Goe:2007,Mor:2012,Mor:2013}. 
Because there are two options to choose between, the choice sequence becomes 
a correlated binary sequence.

 How the subjects' choices are affected by the previous subjects' choices 
 depends on the experimental situation.
 In the canonical setting of an information cascade experiment, 
 there are two urns, $R$ and $B$, which contain red and blue balls  
 in different proportions \cite{And:1997,Kub:2004,Goe:2007}.
 Urn $X\in \{R,B\}$ is fixed at the beginning of each experiment and
 subjects are asked to choose $R$ or $B$.
 If the choice coincides with $X$, the subject receives a  return. 
 Each subject draws a ball from $X$ and obtains some information because
 the color of the ball is usually positively correlated with 
 the type of $X$.
 In addition, he also observes the choices of other subjects' before him.
 If the majority of others' choices suggest $B$, 
 he might choose $B$, even if his ball suggests that $R$ is correct. 
 The same situation continues for the subjects after him, 
 and the majority choice of $B$ tends to continue. This is an information 
 cascade, and it is also called rational herding.

 In an experiment with a two-choice quiz, 
 subjects who did not know the answer to a question
 showed a strong tendency to choose the majority answer \cite{Mor:2012}.
 We call a subject who tends to choose the majority choice a herder. 
 On the other hand, the subject who knows the answer chooses the 
 correct choice independently from others' choices, 
 and  we call him an independent.
 How the herder chooses depends on the return structure.   
 If the return is constant, it is rational to choose the majority choice.
 We call a herder who always chooses the majority choice a digital herder.
 If the return is inversely proportional to the 
 proportion of subjects who have chosen it, as in the parimutuel payoff odds, 
 it is optimal to choose  an option with the same probability as the 
 proportion \cite{Mor:2013}.
 We define the optimal herder an analog herder.   

 The problem to address in studies of information cascades
 is the convergence 
 of the ratio of correct choices $z$ in the long sequence limit 
 \cite{Lee:1993,Goe:2007}. 
 In the voting model where voters are a mixture 
 of independents and analog herders,  
 the model shows a 
 normal-to-superdiffusion phase transition \cite{His:2010}.
 If the voters are a mixture of independents and 
 digital herders,   
 an Ising-like phase transition occurs, and 
 the limit value of 
 the variance of $z$ 
 is the order parameter \cite{His:2011}. 
 If the ratio of herders is low,  
 $z$ converges to a value, and 
 the order parameter
 takes 0. 
 We call the phase a one-peak phase because the distribution of $z$ 
 has only one peak.
 If the ratio is high, 
 $z$ converges to
 one of two values.
 The value to which $z$ converges 
 is determined by a probabilistic process, and 
 the order parameter
 takes a finite and positive value. 
 We call the phase a two-peak phase 
 and the transition between these two phases 
 an information cascade phase transition.
 In the experiment with a two-choice quiz, the ratio
 of herders is controlled by the difficulty of the question. 
 We have reported that the phase transition
 should occur by extrapolating the variance of 
 $z$\cite{Mor:2012}.

 In this paper, we study the critical properties 
 of correlated binary stochastic processes based on  
 finite-size scaling (FSS). 
 In particular, our interest lies in the universality classes 
 of information cascade phase transitions. 
 In equilibrium statistical mechanics, 
 the correlation length plays a key role because it is the unique measure of 
 the collective behavior \cite{Sta:1971}. 
 In the study of the critical behavior of correlated binary 
 sequences, the temporal correlation length 
 $\xi$ should play the same role as 
 in other nonequilibrium phase transitions \cite{Hin:2000,Odo:2004}.
 For the definition of 
 $\xi$, we adopt the second moment 
 correlation time of the autocorrelation 
 function \cite{Car:1995}.
 We study  the scaling properties of $\xi$ and 
 the integrated correlation time $\tau$.  
 We characterize the 
 phases and critical properties of the system by their limit behaviors and 
 critical exponents. 
 We derive some scaling relations and
 estimate the critical exponents, which 
 clarify the universality classes of the information 
 cascade phase transitions.
 In addition, we provide a new data analysis method for
 information cascade experiments.  
 We obtain the 
 finite-size correcting expression for the order parameter.
 Using it, we hope to solve the 
 convergence problem without referring to 
 any specific model.

The paper is organized as follows. 
In Section \ref{sec:fss}, we
define the $r-$th Markov binary process 
$X(t)\in \{0,1\},t\in \{1,2,\cdots\}$, which  
 is a mixture of an independent random variable 
that takes 1 with probability $q$ 
and a random variable that depends on the ratio $z$ 
of the variables taking 1 among recent $r$ variables. 
We consider two models, analog and digital. 
We introduce the FSS
 ansatz for the stochastic process. 
We adopt the second moment correlation time 
of the autocorrelation function $C(t)=\mbox{Cov}(X(1),X(t+1))$ 
as the temporal correlation length $\xi$.
In Section \ref{sec:model_r}, we analyze 
the stochastic process for finite $r$.
$C(t)$ decays exponentially, 
and  there is only one stable renormalization group (RG) fixed point at 
$\lim_{t\to \infty}\xi(t)/t=0$.
We study the $r\to \infty$ limits of the two models 
in Section \ref{sec:model_rinf}. 
$C(t)$ in the analog model obeys a power law, and 
$\lim_{t\to \infty}\xi(t)/t$ is finite.  
Regarding the digital model for $q\neq 1/2$, there are two 
stable RG fixed points corresponding to the  two phases of the 
information cascade phase transition. 
We obtain the scaling relations and  estimate the critical exponents. 
Section \ref{sec:conclusion} presents a summary and 
future problems. 
In the Appendices, we  derive some results in the main text and show
the numerical procedure. 

\section{\label{sec:fss}Models and finite-size scaling analysis }
\subsection{Models}
We define the $r$-th Markov binary processes $X(t)\in \{0,1\},
t\in \{1,2,\cdots,T\}$.  
$X(t+1)$ is a mixture of an independent random 
variable that takes 1 with probability $q$ and a 
random variable
that depends on the ratio $z(t,r)$ of the previous $r$ 
variables $X(s),s\in 
\{t-r+1,t-r+1,\cdots,t\}$, which takes 1.
\begin{equation}
z(t,r) =
\begin{cases}
\frac{1}{r}\sum_{s=t-r+1}^{t}X(s) & \mbox{for}\,\,\,t\ge r,\\
\frac{1}{t}\sum_{s=1}^{t}X(s) & \mbox{for}\,\,\,t<  r .
\end{cases}
\end{equation}
If $t< r$ or in the limit $r\to \infty$, 
$X(t+1)$ depends on all the previous $t$ variables.
It is a simple model for the sequential voting 
process in information cascade experiments, and the two types of 
random variable correspond to the two types of voters: 
 independents and herders.
If voter $t$ is an independent, the probability for the correct choice 
 or the independent variable takes 1 
is $q$.
If voter $t$ is a herder, he obtains information from the previous $r$
subjects by referring to $z(t-1,r)$. If $t=1$, there is no 
available information, 
and the choice is random.
If $t>1$ and under the condition $z(t-1,r)=z$, 
 the probability that herder $t$ chooses the correct option or
the dependent variable takes 1 
is given by the function $f(z)$.
The ratio of the independent and dependent variables is $1-p:p$.
The probability that $X(t)$ takes 1 is then given by 
\begin{equation}
\mbox{Pr}(X(t+1)=1|z(t,r)=z)=(1-p)\cdot q+p\cdot f(z). \label{eq:model}
\end{equation}
The first term comes from the independent random variable with the ratio $1-p$,
and the second term comes from the dependent random variable with 
the ratio $p$. $\mbox{Pr}(X(t+1)=0|z(t,r)=z)$ is 
given as $1-\mbox{Pr}(X(t+1)=1|z(t,r)=z)$.
For the function $f(z)$, we consider two types: 
(i) digital [$\theta(z-1/2)$] and (ii) analog [$z$]. 
Here $\theta(z)$ is a Heaviside function  with 
the convention $\theta(0)=1/2$. 

\subsection{Finite-size scaling ansatz}
We discuss the scaling property of stochastic binary processes 
$\{X(t)\}$. 
The critical behavior
 is governed by a temporal length scale, which we call the 
 correlation length $\xi$. The definition of $\xi$, we adopt the 
 second moment correlation time of 
 the autocorrelation function $C(t)$.
 according to the 
 definition of the second moment correlation length for spin models
\cite{Car:1995}.
$C(t)$ is defined as the covariance 
of $X(1)$ and $X(t+1)$: 
\begin{equation}
C(t)\equiv \mbox{Cov}(X(1),X(t+1))\equiv 
\mbox{E}(X(1)X(t+1))-\mbox{E}(X(1))\mbox{E}(X(t+1)). \label{eq:def_ct}
\end{equation}
Here, the expectation value E$(A)$ of some quantity $A$ is defined 
as the ensemble average over the paths of the stochastic process.
We denote the $n-$th moment of $C(s)$ for the period $s<t$ 
as $M_{n}(t)$. 
\[
M_{n}(t)\equiv \sum_{s=0}^{t-1}C(s)s^{n}.
\]
Hereafter, $t$ in $M_{n}(t)$ plays the role of the time horizon or 
system size of the stochastic process.  The second moment 
correlation time, or correlation length, $\xi(t)$ is defined as
\begin{equation}
\xi(t)\equiv \sqrt{M_{2}(t)/M_{0}(t)}. \label{eq:def_xi}
\end{equation}
 In the study of equilibrium phase transition of spin models, 
 $\xi$ is defined as the logarithm of the ratio of the 
 two lowest eigenvalues by the transfer matirix method. 
 When it is difficult to obtain it, the second moment correlation 
 length is adopted as the proxy.   
\cite{Bar:1983,Bin:1985,Car:1993,Car:1995}.

The integrated correlation time, or relaxation time, 
$\tau(t)$ is defined as
\begin{equation}
\tau(t)\equiv M_{0}(t)/C(0).   \label{eq:def_tau}
\end{equation}
$\tau$ and $\xi$ have the same dimension and one might think that $\tau$
 can play the same role with $\xi$. However,
 the length scale in the critical behaviors of $\xi$ and $\tau$ 
 is $\xi$, not $\tau$ and  the next scaling relation 
 holds for $\xi$.

In FSS, the critical property of any 
long-time observable $A(t)$ for the time horizon $t$ 
 is assumed to be scaled by $\xi(t)/t$. 
 We introduce a scale factor $\sigma$ and 
 the ansatz is written as
\begin{equation}
A(\sigma t)/A(t)=f_{A}(\xi(t)/t) \label{eq:fss},
\end{equation}
which is correct up to terms of order $\xi^{-\omega}$ and $t^{-\omega}$ 
\cite{Car:1995}. 
 Here $f_{A}$ is a universal function and  $\omega$ is a 
correction-to-scaling exponent.
Using the universal function $f_{\xi}$ for $\xi$, we can 
extrapolate $\xi(t)$ for system size $t$ 
to the value for system size $s^{k}t$ as
\begin{equation}
\xi(\sigma^{k}t)=\xi(t)\cdot 
\prod_{l=0}^{k-1}f_{\xi}(\xi(\sigma^{l}t)/\sigma^{l}t).
\label{eq:ext1}
\end{equation}
With this information and $f_{A}$, 
$A(t)$ is extrapolated to $A(\sigma^{n}t)$
as 
\begin{equation}
A(\sigma^{n}t)=A(t)\cdot 
\prod_{k=0}^{n-1}f_{A}\left(
\xi(\sigma^{k}t)/\sigma^{k}t\right).
\label{eq:ext2}
\end{equation}
We extrapolate $A(t)$ 
for a finite time horizon $t$ 
to the limit $t\to \infty$ by Eq.(\ref{eq:ext2}).
We denote the  extrapolated values of $A(t)$
as $A(\infty)$. 

The statistical errors in $A(\infty)$ 
come from three sources: 
(i) the error in $A(t)$, (ii) the error in $\xi(t)$, and (iii) 
the error in $f_{A}$ and $f_{\xi}$ \cite{Car:1995}. 
When a Monte Carlo method is used to obtain
$A(t)$ and $\xi(t)$, 
these errors are estimated as standard errors. 
From them, the errors in $A(\infty)$ are estimated.
In this paper, we integrate the master equation of the system 
and obtain exact
estimates of $A(t)$ and $\xi(t)$.
 There is no statistical error, and only  
 the correction-to-scaling remains in Eq.(\ref{eq:fss}), which 
 will propagate  to $A(\infty)$.
 We estimate the error in $A(\infty)$ as the discrepancy
 between the extrapolated values from different 
 time horizons $t$. If we have the exact value for $A(\infty)$,
 we check the discrepancy between the extrapolated value and 
 the exact value. Please refer to Appendix \ref{sec:Num} 
 for the numerical procedure.

 We make a comment about the choice of $\sigma$.
 It is not crucial if one has the exact universal functions.
 We can even take the limit $\sigma \to 1$ and obtain a differential equation
 for $A(t)$.  
 If one adopt a Monte Carlo method to estimate universal functions 
 by Eq.(\ref{eq:fss}), 
 $\sigma$ should be large enough. Usually and in this paper we take 
 $\sigma=2$\cite{Car:1995}.

\subsection{Three typical cases}
In the following sections, we derive $C(t)$ 
for the stochastic processes [Eq.(\ref{eq:model})] 
and study their scaling  properties. 
In the models, $C(t)$ is estimated using
\begin{equation}
C(t)\equiv \mbox{Cov}(X(1),X(t+1))=p\cdot 
\mbox{Cov}(X(1),f(z(t,r)). \label{eq:Ct} 
\end{equation}
We show that there are three asymptotic 
behaviors of $C(t)$ for the models.
Here, we discuss the scaling properties of $\xi$ and $\tau$ 
for the three cases in advance. 

\begin{enumerate}
\item Exponential decay case

We assume the following functional form for  $C(t)/C(0)$ with some 
positive constants $\delta,p\le 1$:
\begin{equation}
C(t)/C(0)=\delta\cdot p^{t}=\delta \cdot e^{t\log p}.
\end{equation}
$C(t)$ decays exponentially for $p<1$, 
and the system is in the disordered phase.
The expressions 
for $\tau(t)$ and $\xi(t)$ are
\begin{eqnarray}
\tau(t)&=&\delta \cdot \frac{1-p^{t}}{1-p}, \nonumber \\
\xi(t)&=&\sqrt{p \frac{(1-p^{t}-t^{2}p^{t-1}(1-p))
(1-p)+2(p+(t-1)p^{t+1}-tp^{t})}{(1-p)^{2}(1-p^{t})}}. \nonumber 
\end{eqnarray}
In the limit $t \to \infty$, $\tau(t)$ and  
$\xi(t)$ are finite for $p<1$.
\begin{equation}
\lim_{t\to \infty}\tau(t)=\frac{\delta}{1-p} \,\,\,\,\mbox{and}\,\,\,\, 
\lim_{t\to \infty}\xi(t)=\frac{\sqrt{p(1+p)}}{1-p} \label{eq:xi_D}.
\end{equation}
$\xi(t)$ diverges only at $p=1$.
The critical exponent $\nu_{||}$ for 
$\xi \propto |p-p_{c}|^{-\nu_{||}}$ is 
$\nu_{||}=1$, and $p_{c}=1$.

The scaling relation might hold in 
the critical region ($\lim_{t\to \infty}\xi(t)/t>0$), and
we parametrize $p$ as $p=1-\frac{s}{t}$.
With $s$ fixed, we take the limit $t\to \infty$, and $\tau(t)/t$ 
becomes  
\begin{equation}
\lim_{t\to \infty} \tau(t)/t|_{p=1-s/t}
=\frac{\delta}{s}(1-e^{-s})\to \delta \,\,\,\mbox{as}\,\,\, s\to 0 .
\end{equation}
In the same way, we have the limit of $\xi(t)/t$ 
in terms of $s$ as 
\begin{equation}
\lim_{t\to \infty} \xi(t)/t|_{p=1-s/t}
=\frac{1}{s}\sqrt{\frac{2-e^{-s}(2+2s+s^{2})}{(1-e^{-s})}}\to 
\frac{1}{\sqrt{3}}\,\,\mbox{as}\,\,s\to 0.
\end{equation}

We obtain the parametric expressions 
of  the universal functions for $\xi$ and $\tau$ 
in terms of $s$ as follows.
\begin{eqnarray}
f^{D}_{\xi}(s)&\equiv &
\lim_{t\to\infty}
\frac{\xi(2t)|_{p=1-2s/2t}}{\xi(t)|_{p=1-s/t}}=
\sqrt{\frac{(1-e^{-s})}{(1-e^{-2s})}\cdot \frac{2-e^{-2s}(2+4s+4s^{2})}
{2-e^{-s}(2+2s+s^{2})}}\to 2 \,\,\,\mbox{as}\,\,\,s\to 0, \nonumber \\
f^{D}_{\tau}(s)&\equiv&
\lim_{t\to\infty}\frac{\tau(2t)|_{p=1-2s/2t}}{\tau(t)|_{p=1-s/t}}
=
1+e^{-s}\to 2 \,\,\,,\mbox{as}\,\,\,s\to 0. 
\label{eq:universal_D}
\end{eqnarray}
Here, the superscript $D$ indicates ``disordered''. 

We call the scale transformation $t\to 2t$ 
the RG transformation.
Using the extrapolation formulas Eqs.(\ref{eq:ext1}) and (\ref{eq:ext2}), 
because $f^{D}_{\xi}<2$ and $f^{D}_{\tau}<2$ for $s>0$, 
both $\xi(t)/t$ and $\tau(t)/t$ vanish 
in the limit $t\to \infty$, 
as the following relations hold in the RG transformation. 
\[
\xi(2t)/2t<\xi(t)/t\,\,\,\,\mbox{and}\,\,\,\, \tau(2t)/2t<\tau(t)/t .
\]
The RG stable fixed point is characterized by $\lim_{t\to \infty} \xi(t)/t=0$ 
and $\lim_{t\to \infty}\tau(t)/t=0$. It is stable,  
as the infinitesimally perturbed state does return to 
the fixed point under the RG transformation. 

At  $p=1 (s=0)$, both $f_{\xi}$ and $f_{\tau}$ take 2. 
$\xi/t$ and $\tau/t$ are invariant under the RG transformation.
\[
\xi(2t)/2t=\xi(t)/t\,\,\,\,\mbox{and}\,\,\,\, \tau(2t)/2t=\tau(t)/t .
\]
The fixed point is characterized by 
$\lim_{t\to\infty}\xi(t)/t=1/\sqrt{3}$ and $\lim_{t\to\infty}\tau(t)/t=\delta$.
$C(t)$ does not decay, and 
it corresponds to the two-peak phase. 
Further, $\delta$ acts as the order parameter of 
the information cascade transition.
Because $f_{\xi}<2$ for $\xi/t<1/\sqrt{3}$, the infinitesimal 
perturbation to the fixed point 
breaks it. It is an unstable RG fixed point.  

\item Power-law decay case

We assume that $C(t)$ behaves asymptotically with some positive 
constant $\delta,p\le 1$ as  
\[
C(t)/C(0)= \delta\cdot t^{p-1}.
\]
It is easy to estimate $\tau$ and $\xi/t$; the results are:
\begin{equation}
\tau(t)\simeq \frac{\delta}{p}\cdot t^{p}\,\,\,\, \mbox{and}\,\,\,\
\xi(t)/t \simeq \sqrt{\frac{p}{p+2}} \label{eq:xit_SI}. 
\end{equation}
Because $\xi(t)$ is proportional to $t$, 
all states $\xi(t)/t \in [0,1/\sqrt{3}]$ are 
scale invariant because  
$\xi(2t)/2t=\xi(t)/t$ holds. 

The universal functions in terms of $\xi/t$ are
\begin{eqnarray}
f^{SI}_{\xi}(\xi/t)&=&2,  \nonumber \\
\log_{2}f^{SI}_{\tau}(\xi/t)&=&p=\frac{2}{(\xi/t)^{2}-1}
\label{eq:universal_SI}.
\end{eqnarray}
Here, the superscript $SI$ indicates ``scale invariant.''
Because $f^{SI}_{\tau}<2$ for $p<1$, $\lim_{t\to\infty}\tau(t)/t=0$.
At $p=1$, $\lim_{t\to\infty}\tau(t)/t=\delta >0$ and it is the two-peak phase.

\item  Two-peak phase case

In the two-peak phase of 
the information cascade phase transition,
$\lim_{t\to \infty}C(t)>0$. 
We assume the following asymptotic form for $C(t)$
with some positive constant $c$ and a rapidly decaying function  
$d(t)$ as 
\begin{equation}
C(t)/C(0)=c+d(t) \label{eq:c_O}.
\end{equation}
Here, $c$ is defined as
\[
c\equiv \lim_{t\to\infty}C(t)/C(0).
\]
We define $D_{n}(t)$ as $D_{n}(t)=\sum_{s=0}^{t-1}s^{n}d(s)$.  
$\tau(t)$ and $\xi(t)$ are 
\begin{equation}
\tau(t)=c\cdot t+D_{0}(t) \,\,\,\mbox{and}\,\,\,
\xi(t)=\sqrt{\frac{\frac{c}{6}t(t-1)(2t-1)+D_{2}(t)}{ct+D_{0}(t)}}. 
\end{equation}
If we assume $\lim_{t\to \infty}|D_{n}(t)/t^{n+1}|\propto  1/t$,  
we have
\begin{eqnarray}
\tau(t)/t&=&c+\frac{D_{0}(t)}{t}
\to c \,\,\,\mbox{as}\,\,\, t\to \infty, \nonumber \\
\xi(t)/t&=&\frac{1}{\sqrt{3}}
\sqrt{\frac{ct-\frac{3c}{2}+3D_{2}(t)/t^{2}}{ct+D_{0}(t)}}
\to \frac{1}{\sqrt{3}}
 \,\,\,\mbox{as}\,\,\, t\to \infty 
\label{eq:fsc}.
\end{eqnarray}
 These asymptotic behaviors are useful 
 in the estimation of $c$ and 
 $\lim_{t\to\infty}\xi/t$ from the empirical 
 data of correlated binary sequences.
 The universal functions in the two-peak phase
 $f^{TP}_{\xi}\equiv\xi(2t)/\xi(t)$ and $f^{TP}_{\tau}\equiv\tau(2t)/\tau(t)$ 
 are 2
 at $\xi(t)/t=1/\sqrt{3}$, as in the previous two cases at $p=1$. 
 $\xi(t)/t=1/\sqrt{3}$ is invariant under the RG
 transformation. 
 The stability of the RG fixed point 
 depends on the shape of the universal 
 function $f^{TP}_{\xi}(\xi/t)$ for $\xi/t \lesssim 1/\sqrt{3}$.
 $f^{TP}_{\xi}(\xi/t)$ is estimated in the region as
\begin{equation}
f^{TP}_{\xi}(\xi/t)=2+\sqrt{3}(\frac{1}{3}-\frac{\xi}{t}) \label{eq:f_xi_TP}.
\end{equation}
 As $f_{\xi}(\xi/t)> 2$ for $\xi/t \lesssim 1/\sqrt{3}$, the fixed point 
 is RG stable. 
\end{enumerate}

We make three comments. The first is about the relation between the 
order parameter of the information cascade 
phase transition and $c$. In our previous work, we adopted 
the limit value of the variance of the 
ratio $z(t)=\sum_{s=1}^{t}X(s)/t$, which we
denote V$(z(t))$\cite{His:2011}.
If $C(s,s')=C$ for 
$s,s'>>1$, we have 
$\lim_{t\to \infty}\mbox{V}(z(t))
=\lim_{t\to\infty}\sum_{1\le s,s' \le t}C(s,s')/t^{2}=C$. 
Because $\lim_{t\to \infty}C(t)=c\cdot C(0)$, the following relation holds.
\begin{equation}
\lim_{t\to \infty}\mbox{V}(z(t))=c\cdot C(0).\label{eq:VZ_c}
\end{equation}

The second is about 
another representation for $c=\lim_{t\to\infty}C(t)/C(0)$.
We rewrite $C(t)/C(0)$ as the difference between the conditional 
probabilities.
\begin{eqnarray}
C(t)/C(0)&=&
\mbox{Pr}(X(t+1)=1|X(1)=1)-\mbox{Pr}(X(t+1)=1|X(1)=0) \nonumber \\
&=&\mbox{Pr}(X(t+1)=0|X(1)=0)-\mbox{Pr}(X(t+1)=0|X(1)=1).
\label{eq:c_diff}
\end{eqnarray}
If $c=0$, the infinitely departed variable  
$\lim_{t\to\infty}X(t)$ from $X(1)$
does not depend on $X(1)$.  
$c$ represents the strength of the 
information transmission from $X(1)$ 
to $\lim_{t\to\infty}X(t)$.

 The third is about the definition of the second moment 
 correlation length in eq.(\ref{eq:def_xi}).  
 In percolation theory,
 correlation function is defined 
 as the probability that the two sites separated by distance $t$ 
 are in the same cluster\cite{Sta:1991}. 
 In the context, the second moment 
 correlation length represents the typical size of the cluster. 
 For spin models,
 the second moment 
 correlation length for spin-spin correlation function 
 represents the typical 
 size of spin cluster\cite{Bar:1983}. 
 As $C(t)$ is defined 
 as the correlation 
 between $X(1)$ and $X(t+1)$ in eq.(\ref{eq:def_ct}),
 we can regard $\xi(t)$ as the typical size of spin cluster originated 
 from the first voter.

\section{\label{sec:model_r}Models for $r<\infty$}

\subsection{Exact results for $r=1$}
For $r=1$, the analog and digital models are the same, and they are known as 
a correlated random walk \cite{Boh:2000}.
The probability that $X(t+1)$ takes 1 depends on $z(t,1)=X(t)$ as
\begin{equation}
\mbox{Pr}(X(t+1)=1|X(t)=x)=(1-p)\cdot q+p\cdot x \label{eq:r=1}.
\end{equation}
We summarize the results for several quantities 
that are necessary for discussing the scaling relations.
The derivations are given in Appendix \ref{sec:r1}.
The expectation value of 
$X(t)$ is
\begin{equation}
\mbox{E}(X(t))=q+p^{t}(\frac{1}{2}-q) \label{eq:DEXt_r1},
\end{equation}
and the variance of $X(t)$ is $\mbox{V}(X(t))=\mbox{E}(X(t))
\cdot (1-\mbox{E}(X(t)))$.
%\[
%\mbox{V}(X(t))=\mbox{E}(X(t))\cdot (1-\mbox{E}(X(t)))=
%q(1-q)+(q-\frac{1}{2})^{2}(2p^{t}-p^{2t}).
%\]
$X(t)$ exponentially converges to $q$ as $t$ increases.

We put $f(z(t,1))=X(t)$ in Eq.(\ref{eq:Ct}) and 
derive the recursive relation
for $C(t)$.
\[
C(t)=p\cdot \mbox{Cov}(X(1),X(t))=p\cdot C(t-1).
\]
$C(t)=C(0)\cdot p^{t}$, and
the system is in the disordered phase for $p<1$. In the limit $p\to 1$, 
all the variables are completely 
correlated, and the system is in the two-peak  
phase with $c=1$.

The convergence of $\mbox{V}(z(t))$ to zero 
depends on $q$ for $p<1$. 
The convergence exponent $\gamma_{z}$ is defined as 
V$(z(t))\propto t^{-\gamma_{z}}$.
The system is in the one-peak phase if $\gamma_{z}>0$ and in the two-peak phase 
if $\gamma_{z}=0$. 
For $q\neq 1$, V$(z(t))$ obeys the usual 
power law $t^{-1}$ and $\gamma_{z}=1$. 
For $q=1$, V$(z(t))\propto t^{-2}$ and $\gamma_{z}=2$. 
%\[
%\lim_{t\to \infty}V(z(t))\propto
%\begin{cases}
%t^{-1} & q\neq 1 \\
%t^{-2} & q=1.
%\end{cases}
% \]
The system is in the one-peak phase and $c=0$ for $p<1$.
$c$ changes discontinuously  from $c=0$ to $c=1$ at 
$p=1$. The critical exponent $\beta$ for $c\propto |p-p_{c}|^{\beta}$
is zero. 

We comment on the relation between $\gamma_{z}$ and 
the dynamic exponent $Z$. Space and time are different in nature, and 
$Z$ is the scaling exponent between them \cite{Hin:2000}.
The coordinate $S(t)$ for the binary process $\{X(s)\in \{0,1\}\},
s\in \{1,\cdots,t\}$ is defined as $S(t)=\sum_{s=1}^{t}(2X(s)-1)$.  
The variance of $S(t)$ is given as $\mbox{V}(S(t))=
4t^{2}\cdot \mbox{V}(z(t))\propto t^{2-\gamma_{z}} \propto t^{2/Z}$.
The relation between $\gamma_{z}$ and $Z$ is $Z=\frac{2}{2-\gamma_{z}}$.
For normal diffusion, V$(S(t))\propto t$, and $Z=2$. 

%\begin{equation}
%2-\gamma_{z}=2/Z \,\,\,\mbox{or}\,\,\, Z=\frac{2}{2-\gamma_{z}} \label{eq:gamma%_Z}.
%\end{equation}

\subsection{FSS analysis for finite $r$}
\label{sec:FSS}
\begin{figure}[htbp]
\begin{center}
\begin{tabular}{c}
\includegraphics[width=7cm]{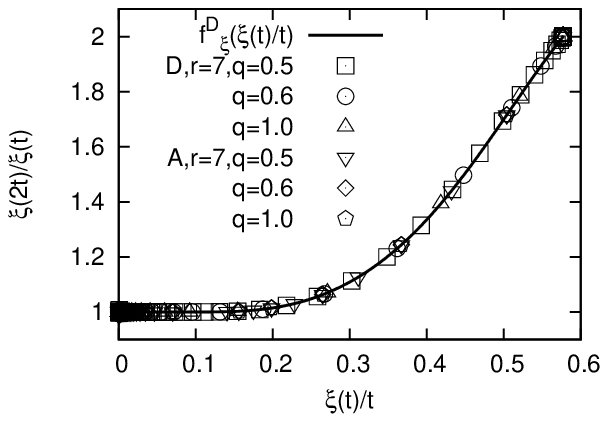}
\vspace*{0.3cm} \\
\includegraphics[width=7cm]{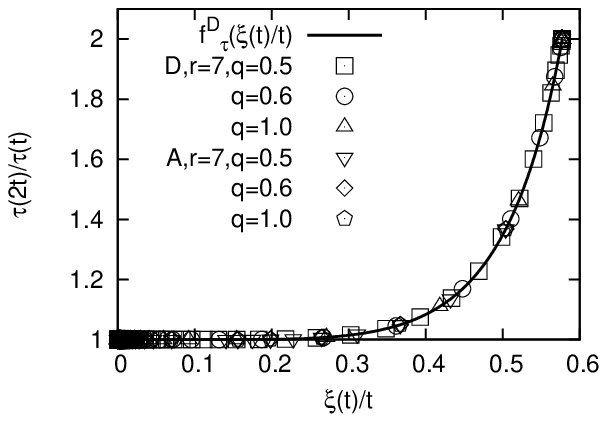} 
 \\
\end{tabular}
\end{center}
\caption{\label{fig:universal_r}
$\xi(2t)/\xi(t)$ and $\tau(2t)/\tau(t)$ versus
$\xi(t)/t$. Symbols indicate different models. 
We set $t=4\times 10^{3}$ and $r=7$.
Curves are universal functions $f^{D}_{\xi}$ 
and $f^{D}_{\tau}$ for the discorded Case, 
which are given in Eq. (\ref{eq:universal_D}).}
\end{figure}

 We study the scaling properties of the models 
for $r=7$ and $q\in \{0.5,0.6,1.0\}$. 
We numerically integrate the models and 
estimate $C(t)$ as in Appendix \ref{sec:Num}. We calculate the ratios 
$\xi(2t)/\xi(t)$ and $\tau(2t)/\tau(t)$ 
for the time horizon $t=4\times 10^{3}$.
We plot them versus $\xi(t)/t$ in Fig. \ref{fig:universal_r}.

All the data for the two models with $r=7$ are on 
the curve of the universal 
functions $f^{D}_{\xi}(\xi/t),f^{D}_{\tau}(\xi/t)$  and
can be used  to estimate $\xi(\infty)$ 
and $\tau(\infty)$. 
The critical behaviors of the models with finite $r$
are the same as in the  exponential decay case. 
Below, we examine the difference between the analog and 
digital models with finite $r$. In particular, we focus on the 
critical behavior of $\xi(\infty)$ for $p\lesssim p_{c}=1$.

\subsection{Analog model with finite $r$}
%We recall the probabilistic rule of the model.
%\[
%\mbox{Pr}(X(t)=1|z(t-1,r)=z)=(1-p)\cdot q+p\cdot z.
%\]
%\color{red}
We put $f(z(t,r))=z(t,r)$ in Eq.(\ref{eq:Ct}).
For $t\le r $, $z(t,r)=z(t)=\frac{t-1}{t}z(t-1)+\frac{1}{t}X(t)$.
Because  $X(t)\propto p\cdot z(t-1)$, we have the recursive relation.
\[
C(t)=\frac{t-1+p}{t}\cdot C(t-1)\,\,\,\,\mbox{for}\,\,\, t\le r.
\]
We solve the recursive relation and obtain
\begin{equation}
C(t)/C(0)=\prod_{s=1}^{t}\frac{s-1+p}{s}\propto t^{p-1}\label{eq:Ct_A}.
\end{equation}

For $t\ge r+1$, $z(t,r)=z(t-1,r)+\frac{1}{r}(X(t)-X(t-r))$.
Because $X(t-r)\propto p\cdot z(t-r-1,r)$, we have the next recursive relation.
\[
C(t)-C(t-1)=\frac{p}{r}\cdot (C(t-1)-C(t-r-1))\,\,\,\, \mbox{for}\,\,\,\, t\ge r+1.
\]
On the basis of the results of FSS analysis, 
we assume exponential decay for $C(t)\propto p_{r}^{t}$.
We see that $p_{r}$ obeys the following relation.
\begin{equation}
p_{r}^{r}(p_{r}-1)=\frac{p}{r}(p_{r}^{r}-1) \label{eq:pr_A}.
\end{equation}
In the critical region where $p=1-\epsilon$ for 
some small positive number $\epsilon>0$,  $p_{r}$ 
can be written as $p_{r}=1-\frac{\epsilon}{r}$, and 
$\lim_{r\to\infty}p_{r}=1$.
Because $\xi(\infty)\propto 1/(1-p)=1/\epsilon$ for 
the exponential decay case,
$\xi(\infty)$ in the analog model with finite $r$ behaves as $
\xi(\infty)\propto \frac{r}{\epsilon}$.
$\xi(\infty)$ diverges linearly 
with $r$. $\nu_{||}$ is 1 and 
does not depend on $r$.

\subsection{Digital model with finite $r$}
%We study the digital model with finite $r$. 
We map the digital model $(r,q,p)$
to the model with $(r=1,q_{r},p_{r})$ \cite{His:2014}.
We recall the probabilistic rule of the model.
\[
\mbox{Pr}(X(t)=1|z(t-1,r)=z)=(1-p)\cdot q+p\cdot \theta(z-1/2).
\]
To facilitate the treatment, we modify 
the model so that 
$X(s),s\in \{r(t-1)+1,\cdots,rt\}$ obeys the following rule
for $t\ge 1$.
\[
\mbox{Pr}(X(s)=1|z(r(t-1),r)=z)=(1-p)\cdot q+p\cdot \theta(z-1/2).
\]
All the $r$ variables $X(s),s\in \{r(t-1)+1,\cdots,rt\}$ are affected by 
the same $z(r(t-1),r)$.
We group the $r$ variables in a new variable
$X_{r}(t)$ through $z(rt,r)$ by the relation 
\begin{equation}
X_{r}(t)=\theta(z(rt,r)-1/2)=\theta\left(\frac{1}{r}\sum_{s=1}^{r}X(r(t-1)+s)
-1/2\right).
\end{equation}
This is a real space renormalization transformation \cite{Sta:1995} and 
$X_{r}(t)$ depends only on the previous $X_{r}(t-1)$. We write the 
probabilistic rule for $X_{r}(t)$ as
\[
\mbox{Pr}(X_{r}(t)=1|X_{r}(t-1)=x)=(1-p_{r})\cdot q_{r}+p_{r}\cdot x.
\]
If $X_{r}(t-1)=0$, the probability that $X_{r}(t)$ takes 1 is 
$(1-p_{r})q$. In the modified dynamics of $X(s)$, the probability 
is given as
\begin{equation}
(1-p_{r})q_{r}=\pi_{r}((1-p)q). \label{eq:Xr1}
\end{equation}
Here, $\pi_{r}(x)$ is defined as 
$\pi_{r}(x)=\sum_{n=(r+1)/2}^{r}{}_{r}C_{n}\cdot x^{n}(1-x)^{r-n}$.
Likewise, we estimate $(1-p_{r})(1-q_{r})$ as $\pi_{r}((1-p)(1-q))$,
 and we have the following explicit expressions for $q_{r},p_{r}$:
\begin{eqnarray}
q_{r}&=&\frac{\pi_{r}((1-p)q)}{\pi_{r}((1-p)q)+\pi_{r}((1-p)(1-q))}, \\
p_{r}&=&1-(\pi_{r}((1-p)q)+\pi_{r}((1-p)(1-q))) \label{eq:Xr2}.
\end{eqnarray}
For large $r$, $\pi_{r}(x)$ behaves as
\begin{equation}
\lim_{r\to\infty}\pi_{r}(x)\simeq 
\begin{cases}
1 & x>1/2, \\
1/2 & x=1/2, \\
0 & x<1/2. \\
\end{cases}
\end{equation}
We study the transformation $(q,p)\to (q_{r},p_{r})$, which has five fixed points. 

\begin{figure}[htbp]
\begin{center}
\includegraphics[width=6cm]{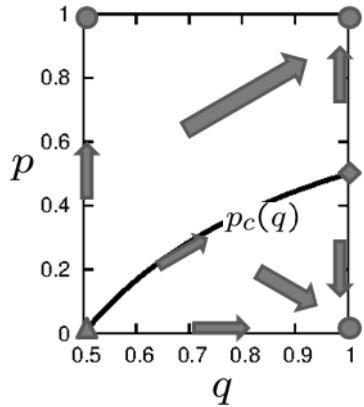}
\end{center}
\caption{\label{fig:RG}
Plot of transformation $(q,p)\to (q_{r},p_{r})$.
Arrows indicates the direction of movement from $(q,p)$ to 
$(q_{r},p_{r})$.
There are three stable fixed points at $(1,1),(1,0),(1/2,1)$ (filled circles) 
and two unstable fixed points at $(1/2,0),(1,1/2)$ (filled triangle and 
diamond, respectively). 
}
\end{figure}

When $q=1/2$, 
$\pi_{r}((1-p)q)=\pi_{r}((1-p)(1-q))$ and 
$q_{r}=1/2$.
For $p>0$,
$\lim_{r\to \infty}\pi_{r}((1-p)q)=\lim_{r\to \infty}\pi_{r}((1-p)(1-q))
=0$ because $(1-p)q<1/2$ and $(1-p)(1-q)<1/2$. We have $\lim_{r\to \infty}p_{r}=1$.
If $p=0$, $\pi_{r}(q)=\pi_{r}(1-q)=1/2$, and we have $p_{r}=0$.
The fixed points of the transformation are $(1/2,0)$ and $(1/2,1)$ 
The former fixed point is unstable, and the latter is stable 
on the $p$ axis. 

When $q>1/2$, $(q,p)$ on the line $p(q)=1-1/2q$ 
moves along it under the transformation because 
$(q_{r},p_{r})$ also satisfies $p_{r}=1-1/2q_{r}$.
We call this line the critical line and denote $p_{c}(q)=1-1/2q$.
Because $(1-p)q=1/2$ and $(1-p)(1-q)<1/2$ on the critical line, we have  
$\lim_{r\to \infty}(q_{r},p_{r})=(1,1/2)$.
If $p<p_{c}(q)$, because $(1-p)q>1/2$ and $(1-p)(1-q)<1/2$,
we have $\lim_{r\to\infty}(q_{r},p_{r})=(1,0)$. 
$(1,0)$ is a stable fixed point.  
In the region $p>p_{c}(q)$,
$(1,1)$ is also a stable fixed point. 
To study the critical region $p\lesssim 1$, we 
write $p=1-\epsilon$. $\pi_{r}(\epsilon q)
\propto (\epsilon q)^{(r+1)/2}$, and $\pi_{r}(\epsilon (1-q))\propto 
(\epsilon (1-q))^{(r+1)/2}$. With these expressions, we obtain 
$q_{r}\simeq q^{(r+1)/2}/(q^{(r+1)/2}+(1-q)^{(r+1)/2})$ and 
$1-p_{r}\propto \epsilon^{(r+1)/2}$.
In the limit $r\to \infty$, 
$(q_{r},p_{r})$ converges to $(1,1)$, which suggests  that
$(1,1)$ is stable under the transformation.
$\nu_{||}$ is estimated as $(r+1)/2$.

We summarize the results in Fig.\ref{fig:RG}.
Under the transformation, 
$(1,1),(1,0)$ and $(1/2,1)$ are stable, and $(1/2,0)$ is unstable.
$(1,1/2)$ has one stable and one unstable direction.
The critical properties of the digital model with finite $r$ 
are almost the same as those of the disordered case
except for $\nu_{||}=(r+1)/2$.

\subsection{Critical behavior of $\xi(\infty)$ versus $r$}
\begin{figure}[htbp]
\begin{center}
\includegraphics[width=8cm]{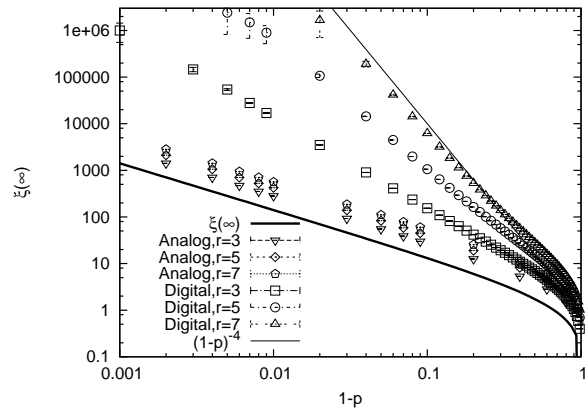}
\end{center}
\caption{\label{fig:Xi_r_AD}
$\xi(\infty)$ versus $1-p$. 
$\xi(\infty)$ is the extrapolated value of $\xi(t)$ for the analog  
and digital models
with $t=4\times 10^{3}$, $q=0.6$, and $r\in\{3,5,7\}$.
For $r=1$, we plot Eq.(\ref{eq:xi_D}) with a thick solid line.
 The thin solid line represents $(1-p)^{-4}$ for comparison
 with the digital model with $r=7$.
The error bars are the absolute values of the difference 
in the extrapolated values from different time horizons, 
$4\times 10^{3}$ and $2\times 10^{3}$. 
}
\end{figure}	
We compare the dependence of the critical behavior 
of $\xi(\infty)$ on $r$ in the two models. 
We show a double logarithmic plot of $\xi(\infty)$ 
versus $1-p$ for $r\in \{3,5,7\}$ in Fig. \ref{fig:Xi_r_AD}.
When $\xi(\infty)$ approaches $10^{6}$, the error bars 
become large, and one sees that 
$t=4\times 10^{3}$  is not sufficient to exceed $\xi=10^{6}$. 
For $r=1$, $\xi(\infty)$ is given by Eq.(\ref{eq:xi_D}),
and it diverges as $\propto (1-p)^{-1}$.
In the analog model with $r$, $\nu_{||}=1$.
In the digital model with $r$, $\nu_{||}=(1+r)/2$.
 For $r=7$, $\nu_{||}=4$, which is consistent with the 
absolute value of the slope in the figure.  
 
\section{\label{sec:model_rinf}$r\to \infty$ limits of models}
In the previous section, we showed that $p_{r}$ approaches $1$
with $r$ in the critical region $p\lesssim 1$
and that $\xi(\infty)$ diverges in the limit $r\to \infty$ 
for both the analog and digital models. 
However, the nature of the divergence differs greatly.
In the analog model, $\xi(\infty)$ 
diverges linearly with $r$. 
In the digital model, $\xi(\infty)\propto (1-p)^{-(r+1)/2}\propto 
e^{r(-\log(1-p)/2)}$, and $\xi(\infty)$ diverges exponentially with $r$. 
In this section, we show that 
the difference leads to completely different critical 
behaviors in the limit $r\to \infty$.

\subsection{Analog model in the limit $r\to\infty$}
We recall some results 
for the analog model in the 
limit $r\to\infty$ \cite{Hod:2004,Hui:2008,His:2010}. 
The model shows the normal-to-superdiffusion 
phase transition, if $q\neq 1$.
For $p>p_{s}=1/2$, $\gamma_{z}=2-2p<1$, and the system is in the 
superdiffusion phase.
If $p<p_{s}=1/2$, $\gamma_{z}=1$, and it is in the normal diffusion phase; further, 
at $p=p_{s}$, there is a logarithmic correction to the scaling of $V(z(t))$.
If $q=1$, $\gamma_{z}=2-2p$, and $\gamma$ 
changes continuously with $p$;  
the phase transition does not occur.
The two-peak phase of the information cascade phase transition 
exists at $p=1$, and for $p<1$, it is in the one-peak phase.

The results are summarized as
\begin{equation}
\lim_{t\to \infty}V(z(t))\propto
\begin{cases}
t^{-1} & p<p_{s}=1/2,q\neq 1, \\
t^{-1}\log t & p=p_{s},q\neq 1, \\
t^{2p-2} & p>p_{s},q\neq 1, \\
t^{2p-2} & q=1 \\
1/4 & p=1
 \label{eq:VZ_A} 
\end{cases}
\end{equation}

When we take the limit $r\to \infty$ in
Eq.(\ref{eq:Ct_A}), $C(t)$ shows a power law decay 
$C(t) \sim t^{p-1}$
for large $t$.
The system is scale invariant, and  
the scaling relations for $\xi$ and $\tau$ are given 
in Eq. (\ref{eq:universal_SI}).  

\begin{figure}[htbp]
\begin{center}
\begin{tabular}{c}
\includegraphics[width=7cm]{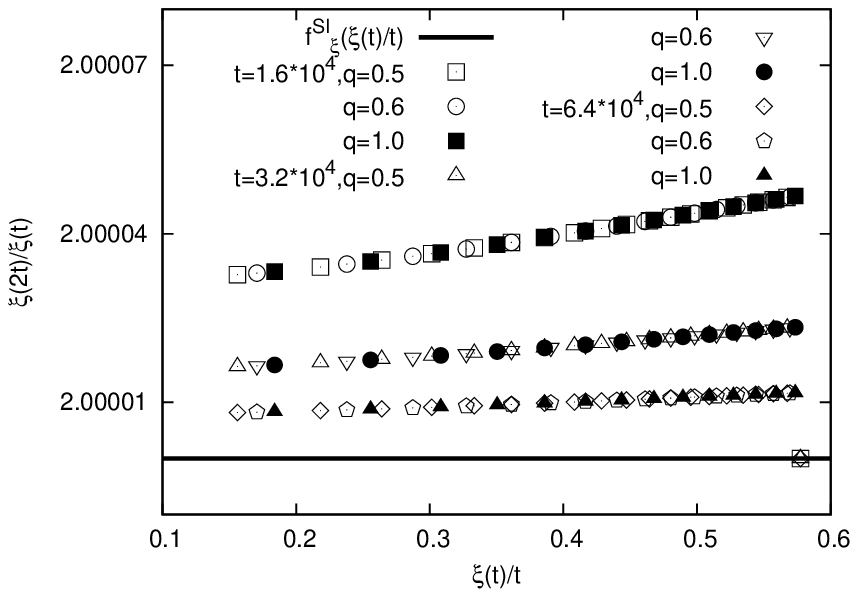}
\vspace*{0.3cm} \\
\includegraphics[width=7cm]{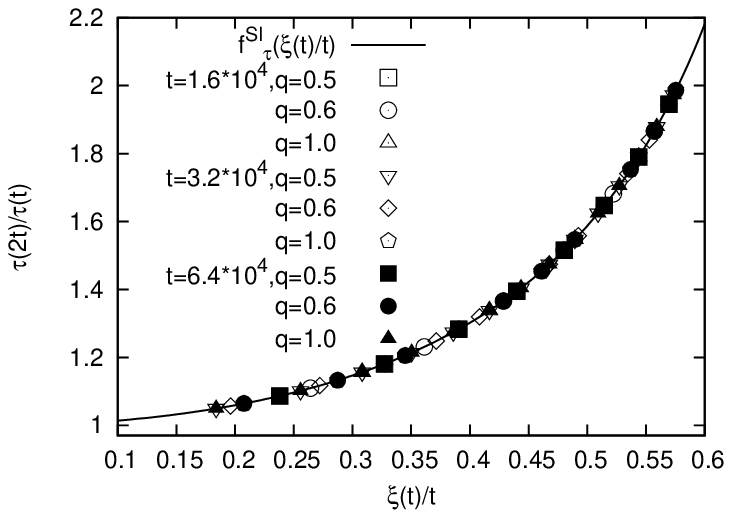} 
 \\
\end{tabular}
\end{center}
\caption{\label{fig:universal_A_rinf} 
$\xi(2t)/\xi(t)$ and $\tau(2t)/\tau(t)$ versus
$\xi(t)/t$ for the analog model in the limit $r\to\infty$. 
Symbols and lines indicate different time horizons $t$ or $q$ or
universal functions $f^{SI}_{\xi},f_{\tau}^{SI}$ in Eq.(\ref{eq:universal_SI}).
We adopt $t\in \{1.6\times 10^{4},3.2\times 10^{4},6.4\times 10^{4}\},
q\in \{0.5,0.5,1.0\}$}.
\end{figure}

\subsection{FSS analysis of the analog model 
in the limit $r\to \infty$}
\label{subsec:fss_a}
We calculate the ratios of $\xi$ and $\tau$
for the time horizon $t \in \{1.6\times 10^{4},
3.2\times 10^{4},6.4\times 10^{4}\}$ 
and $q\in \{0.5,0.6,1.0\}$.
We plot the ratios as functions of  $\xi(t)/t$ in Fig.
\ref{fig:universal_A_rinf}.
The top panel shows $\xi(2t)/\xi(t)$ versus $\xi(t)/t$.
The ratio is almost two for any $\xi(t)/t$, and  
only a small 
correction-to-scaling to $f^{SI}_{\xi}=2$ appears
up to terms of order $1/t$.
The bottom panel shows $\tau(2t)/\tau(t)$ versus $\xi(t)/t$,
 and the ratios are on the curve of $f^{SI}_{\tau}$ in
Eq.(\ref{eq:universal_SI}).

We comment on the relation between 
$C(t)=C(1,t+1)\propto t^{p-1}$ and the behavior of V$(z(t))$  in 
Eq.(\ref{eq:VZ_A}).
If $C(t,t')$ is assumed to depend on $t,t'$ through the difference
$|t-t'|$ as $C(t,t')\propto |t-t'|^{p-1}$, V$(z(t))$ 
behaves as $t^{p-1}$.
\[
\mbox{V}(z(t))=
\sum_{1\le s,s'\le t}C(s,s')/t^{2}\sim \int^{t}(t-s)\cdot s^{p-1}ds/t^{2}
\sim t^{p-1}.
\]
We have $\gamma_{z}=p-1$.
The results in Eq.(\ref{eq:VZ_A}) suggest that the assumption
of the translational invariance 
is wrong.
$C(t,t')$ has a more complex structure, and  
we study the relation between  $C(t,t')$
 and V$(z(t))$ in Appendix \ref{sec:A_rinf}.

\subsection{Digital model in the limit $r\to\infty$}
We recall some exact results for the digital model
in the limit $r\to \infty$ \cite{His:2011}. 
The model shows 
the phase transition between the one-peak and two-peak phases 
as $p$ passes the critical value $p_{c}(q)=1-1/2q$.
Note that $p_{c}(q)$ is the same as the critical line 
for the digital model with finite $r$.
In the one-peak phase, the probability distribution of 
$z(t)$ has a peak at $z=z_{+}\equiv (1-p)q+p$ in the limit $t\to \infty$.
In the two-peak phase, there are two peaks, at $z=z_{-}\equiv (1-p)q$ and 
$z_{+}$. The probability that $z(t)$ converges to the 
lower peak at $z_{-}$ is a continuous function of $p$ and 
takes a positive value for $p>p_{c}$.
The  derivative of the probability at $p_{c}$ becomes discontinuous
in the limit $t\to \infty$, 
which indicates  that
the phase transition for $q>1/2$ is a continuous nonequilibrium 
phase transition. 

\subsection{FSS analysis of the digital model in the limit $r\to \infty$}
\begin{figure}[htbp]
\begin{center}
\begin{tabular}{c}
\includegraphics[width=7cm]{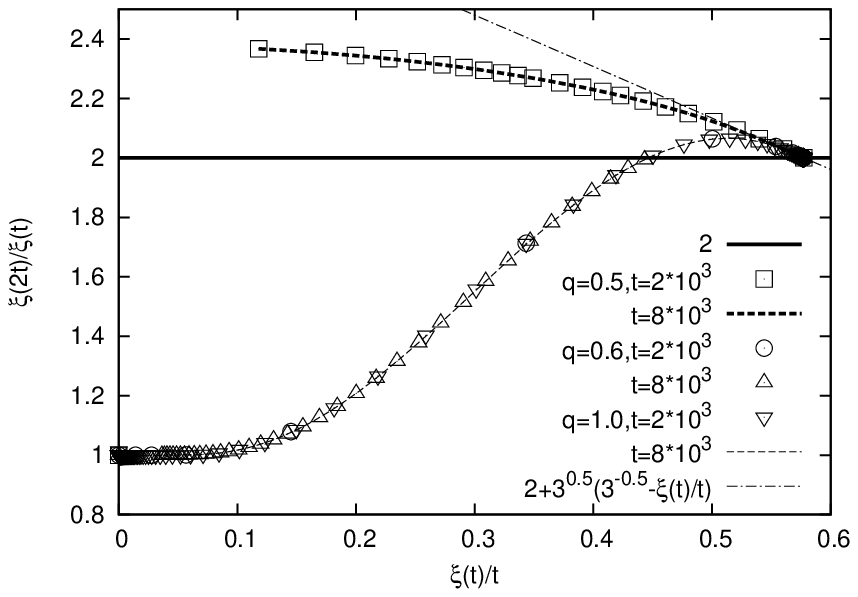}
\vspace*{0.3cm} \\
\includegraphics[width=7cm]{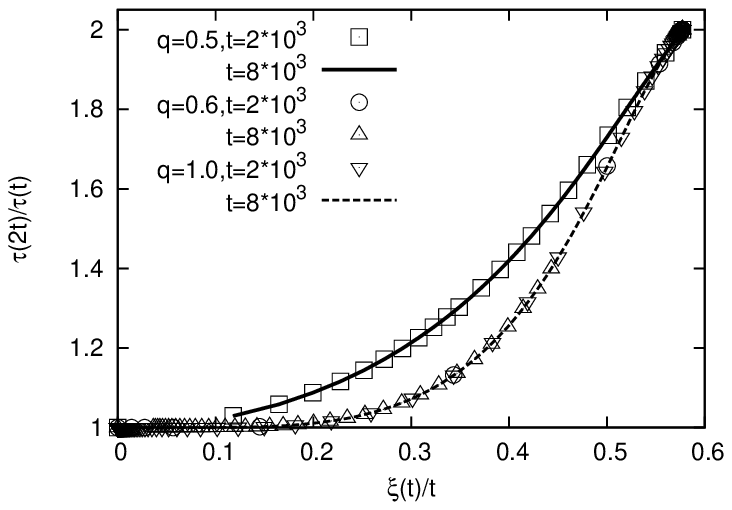} 
 \\
\end{tabular}
\end{center}
\caption{\label{fig:universal_D_rinf} 
$\xi(2t)/\xi(t)$ and $\tau(2t)/\tau(t)$ versus
$\xi(t)/t$ for the digital model in the limit $r\to \infty$. 
Symbols and lines indicate different time horizons $t$ or $q$. 
We adopt $t\in \{2\times 10^{3},8\times 10^{3}\}$ and 
$q\in \{0.5,0.6,1.0\}$.
We also plot Eq.(\ref{eq:f_xi_TP}) in the upper figure.}
\end{figure}	

We calculate the ratios of $\xi$ and $\tau$ 
for the time horizons $t\in \{2\times 10^{3},8\times 10^{3}\}$
 and $q\in \{0.5,0.6,1.0\}$
and  plot the results in Fig.\ref{fig:universal_D_rinf}. 
The scaling relation 
for $\xi(t)$ indicates that there are 
two stable RG fixed points at $\xi/t=0$ and $\xi/t=1/\sqrt{3}$ for $q> 1/2$.
Further, $p_{c}$ is determined by $\xi(2t)/\xi(t)=2$, and 
the system is in the two-peak (one-peak) phase for $p>p_{c} (p<p_{c})$.
We determine $p_{c}$ with the numerical data for $t=8\times 10^{3}$
 and recover the exact result $p_{c}=1-1/2q$ 
to an accuracy of 1\%. For $p>p_{c}$,
$\lim_{t\to\infty}\xi(t)/t=1/\sqrt{3}$.
For $p<p_{c}$, because $\xi(2t)/\xi(t)<2$, $\lim_{t\to\infty}\xi(t)/t=0$, and $
\lim_{t\to \infty}\xi(t)$ is finite. 
At $p=p_{c}$, $\xi(2t)/\xi(t)=2$ and  
the system is scale invariant.
We find $\tau(2t)/\tau(t)=2^{1/2}$ and 
$\xi(t)/t=1/\sqrt{5}$ at $\xi(2t)/\xi(t)=2$.
For $q=1/2$, there is only one stable RG 
 fixed point at $\xi/t=1/\sqrt{3}$. For $p>0$, the system is 
 in the two-peak phase, and the fixed point at $\xi/t=0$ 
is unstable.
 
From the scaling relation for $\tau$ in the bottom panel, 
for $p>p_{c}(q)$, $\tau(2t)/\tau(t)$ converges to 2 in the limit 
$t\to \infty$; 
$\lim_{t\to \infty}\tau(t)/t$ takes a positive value, and it is $c$ 
in Eq.(\ref{eq:fsc}). In the two-peak phase, 
$\lim_{t\to\infty}\xi(t)/t=1/\sqrt{3}$, and $\lim_{t\to\infty}\tau(t)/t=c>0$.
For $p<p_{c}$, $\tau(\infty)$ is finite, and $\lim_{t\to\infty}\tau(t)/t=0$. 
If $q=1/2$, $\lim_{t\to\infty}\xi(t)/t=1/\sqrt{3}$ for $p>0$, and 
$\lim_{t\to \infty}\tau(2t)/\tau(t)=2$. 
We find $\lim_{t\to\infty}\tau(t)/t=c>0$ for $p>0$.

\subsection{Scaling analysis of $C(t)/C(0)$}
\begin{figure}[htbp]
\begin{center}
\includegraphics[width=8cm]{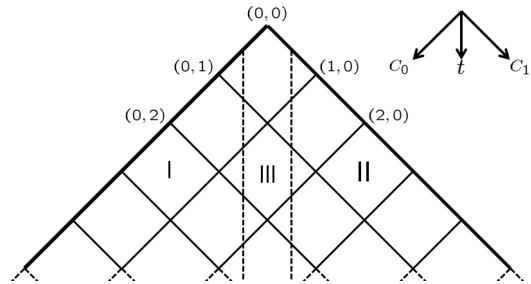}
\end{center}
\caption{\label{fig:dp}
Directed path representation of $\{X(t)\}$ as $(C_{1},C_{0})=(m,n)$.
}
\end{figure}	
We estimate $C(t)/C(0)$ for $q=1$ using our previous 
result \cite{His:2011}.
We introduce a tilted two-dimensional square lattice  
$(m,n),m,n\in \{0,1\cdots \}$ with the origin at the top 
of it. 
We describe the stochastic process  
as a directed path  on the lattice (Fig.\ref{fig:dp}). 
We map $X(t),t\in \{1,2,\cdots\}$ 
to a path on the lattice as $(m,n)=(C_{1},C_{0})$ with $C_{1}=
\sum_{s=1}^{t}X(s)$ and $C_{0}=t-C_{1}$.
The path starts from the top [$(0,0)$], and 
if $X(t)=1 (0)$, it moves to the lower right (left).
We refer to the regions $m<n,m>n$, and $m=n$ as
I,II, and III, respectively. 
The probabilistic rules 
are summarized as
\begin{equation}
\mbox{Pr}(X(t)=0)=
\begin{cases}
(1-p)(1-q)+p\equiv A & m<n,I,\\
(1-p)(1-q)+p/2\equiv B & m=n,III, \\
(1-p)(1-q)\equiv C & m>n,II .
\label{eq:p_D2} 
\end{cases}
\end{equation}

Now we set $q=1$, and the wall $n=m-1$ in II becomes an 
absorbing wall for the path. If a path enters II, 
 it cannot return to III. To estimate $C(t)/C(0)$
 in Eq.(\ref{eq:c_diff}), one needs to know 
$\mbox{Pr}(X(t+1)=0|X(1)=0)$ as $\mbox{Pr}(X(t+1)=0|X(1)=1)=0$ for $q=1$. 
 We denote the number of paths 
 from $(0,0)$ to $(s,s)$ in I and III 
that enter III $k$ times as 
 $A_{s}^{k}=k(2s-k-1)!/s!(s-k)!$ \cite{His:2011}.
 The probability that a  path starts from $(0,1)$ and 
 reaches $(s,s)$  is then given as
\begin{equation}
\mbox{Pr}(z(t=2s)=1/2|X(1)=0)=
\sum_{k=1}^{s}A_{s}^{k}\cdot p^{s-1}(1-p)^{s}/2^{k-1} \label{eq:z=1/2}.
\end{equation}
We denote the probability that a path starts from $(0,1)$ and 
stays in region $J \in \{\mbox{I,II,III}\}$ at $t$ as $P_{J}(t|X(1)=0)$.
For large $t$, the $k=1$ term in Eq.(\ref{eq:z=1/2}) is 
dominant, we have
\begin{equation}
P_{\mbox{III}}(t=2s|X(1)=0)\simeq \frac{2}{p}\cdot 
\frac{e}{8\sqrt{\pi}} s^{-3/2}(4p(1-p))^{s} 
\,\,\,\mbox{for}\,\,\,t>>1. 
\label{eq:III} 
\end{equation}
The probability that a path enters II from 
$(s,s)$ to $(s+1,s)$ is $(1-p/2)$, 
and we have the following expression.
\begin{equation}
P_{\mbox{II}}(t=2s+1|X(1)=0)=\sum_{u=1}^{s}P_{\mbox{III}}(2u|X(1)=0)
\cdot (1-p/2).
\end{equation}
Because $P_{I}(t|X(1)=0)\simeq 1-P_{II}(t|X(1)=0)$ for large $t$ and 
$\mbox{Pr}(X(t+1)=0)=p$ for $z(t)<1/2$, we have
\[
C(t)/C(0)= 
p\cdot P_{\mbox{I}}(t|X(1)=0)=
p\cdot (1-P_{\mbox{II}}(t|X(1)=0))
\,\,\,\mbox{for}\,\,\,t>>1. 
\]
The limit value $\lim_{t\to\infty}C(t)/C(0)$, which we denote as 
$c(q=1,p)$, is then estimated using the results 
in \cite{His:2011} as 
\[
c(q=1,p)=\lim_{t\to\infty}C(t)/C(0)=
\begin{cases}
0 & p\le 1/2,\\
\frac{4p-2}{(1+p)} & p\ge 1/2. 
\end{cases}
\]
Using $c(q=1,p)$ we rewrite the expression for $C(t)/C(0)$:
\begin{eqnarray}
C(t=2s)/C(0)=c(1,p)+p\cdot 
\sum_{u=s+1}^{\infty}P_{\mbox{III}}(2u|X(1)=0)\cdot (1-p/2). 
\end{eqnarray}
Putting the asymptotic form Eq.(\ref{eq:III}) 
for $P_{III}(2u|X(1)=0)$ in the expression,
we obtain
\[
C(t)/C(0)\sim c(1,p)+p\cdot \int_{t}^{\infty}ds 
\frac{e}{8\sqrt{\pi}}s^{-3/2}e^{-s/\xi(p)}\cdot (1-p/2).
\]
Here, we define $\xi(p)=-/\log\sqrt{4p(1-p)}$.
Using the results, we estimate the critical exponents.
We expand $c(1,p)$ around $p=p_{c}(1)=1/2$; then 
we have $\beta=1$ for $C(1,p)\propto |p-1/2|^{\beta}$.
For $\nu_{||}$, we expand $\xi(p)$ around $p_{c}=1/2$ and obtain $\nu_{||}=2$.
At $p=p_{c}$, $C(t)$ behaves as $t^{-\alpha}$ with $\alpha=1/2$ as
$\xi(p)$ diverges.

The FSS results suggest that the phase transition for 
$q>1/2$ is governed by the fixed point $(1,1/2)$
of the transformation $(q,p)\to (q_{r},p_{r})$.
In particular, at $p=p_{c}(q)$, $\mbox{Pr}(X(t)=0)=1/2$ in I 
$(m<n)$, and it is a simple symmetric random walk.
In II and III, $\mbox{Pr}(X(t)=1)>1/2$, and 
it is not symmetric.
The nonrecurrence probability that a simple 
random walk does not return to 
the diagonal up to 
$t$ behaves as $t^{-1/2}$, which can be checked by
Eq.(\ref{eq:z=1/2}).  
$C(t)$ is proportional to the probability
that the random walker remains in I because 
it is difficult for the random walker
to return from II to I even for $q<1$.
Thus, $C(t)\propto P_{I}(t|X(1)=0)
\sim t^{-1/2}$ holds generally.
Furthermore, because the paths of the random walker in I are concentrated 
around $z\sim 1/2$, the same asymptotic behaviors 
$\mbox{Pr}(X(t')=0|X(t)=0)\sim |t'-t|^{-1/2}$ should hold for $t,t'>0$.
This suggests that the translational invariance is recovered 
at $p=p_{c}$ and $C(t,t')\propto |t-t'|^{-1/2}$.  
From the discussion in \ref{subsec:fss_a}, 
we have $\gamma_{z}=1/2$. 
However, there are many assumptions in the 
discussion, 
so $\gamma_{z}=1/2$ should be studied numerically.
 
The above directed path picture suggests the relation between the 
information cascade phase transition and a phase transition 
to absorbing states. For $q=1$, the 
wall $n=m-1$ in II becomes  an  absorbing wall, and 
a path does not return to I if it touches the wall.
Region II 
becomes an absorbing state for the path.   
$C(t)$ is the survival probability of 
the path in I, which is the order parameter of 
the phase transition into absorbing states \cite{Hin:2000}. 
We can assume the following scaling form for $C(t)$ 
in the critical region for $q>1/2$.
\begin{equation}
C(t)/C(0)\propto t^{-\alpha}\cdot g(t/\xi(t)),\,\,\,\mbox{and}\,\,\,\alpha=1/2.
\label{eq:p_D} 
\end{equation}
Here, we assume that $C(t)$ is scaled by 
$\xi(t)$ with a universal scaling function $g$. 
With this expression, we can 
derive the relation between $\tau(\infty)$ and 
$\xi(\infty)$ as $\tau(\infty)\propto 
\xi(\infty)^{1-\alpha}$ for $p\lesssim p_{c}(q)$. 
For $p\lesssim p_{c}(q)$, $\xi(\infty)$ should diverge as 
$|p-p_{c}(q)|^{-\nu_{||}}$, and we have the relation 
$\nu_{\tau}=(1-\alpha)\nu_{||}$ for 
$\nu_{\tau}$ of $\tau(\infty)\propto |p-p_{c}(q)|^{-\nu_{\tau}}$.
In the limit $t\to \infty$ for $p\gtrsim p_{c}(q)$, 
$\lim_{t\to\infty}C(t)/C(0)=c>0$.
 Because $g(x)$ should behave as $g(x)\sim x^{\alpha}$
to cancel $t^{-\alpha}$, 
we have
\[
c\propto \xi^{-\alpha}\propto |p-p_{c}(q)|^{\alpha\nu_{||}}.
\]
We have the scaling relation $\beta=\nu_{||}\cdot \alpha$. 

We can also estimate $\nu_{||}$ by the 
relation $p_{r}$ in Eq.(\ref{eq:Xr2}).
Because $(p,q)=(1/2,1)$ is a fixed point under 
the transformation $X\to X_{r}$,   
the following relation holds near the fixed point. 
\[
\lim_{r\to \infty}
\frac{1}{r}|p-1/2|^{-\nu_{||}}=
\lim_{r\to \infty}|p_{r}-1/2|^{\nu_{||}}.
\]
With Eq. (\ref{eq:Xr2}), we have
\[
(p_{r}-1/2)\simeq \sqrt{\frac{2r-2}{\pi e^{2}}}(p-1/2).
\]
Then we can estimate $\nu_{||}$ as
\[
\nu_{||}=\lim_{r\to \infty}1/\log_{r} \sqrt{\frac{2r-2}{\pi e^{2}}}=2.
\]
The result, $\beta=1,\alpha=1/2$ and $\nu_{||}=2$, 
is consistent with the scaling 
relation $\beta=\nu_{||}\cdot \alpha$.
These scaling relations suggest that only two exponents, $\beta$ and 
$\nu_{||}$, might be sufficient to characterize
the universality class of the information cascade 
phase transition.

\subsection{Critical behaviors}

\begin{figure}[htbp]
\begin{center}
\begin{tabular}{cc}
\includegraphics[width=6cm]{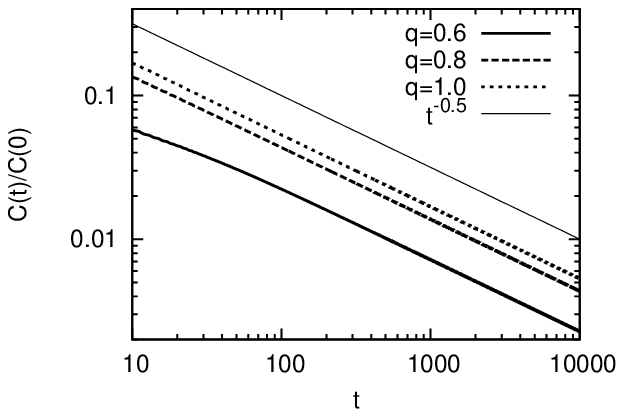} &
\includegraphics[width=6cm]{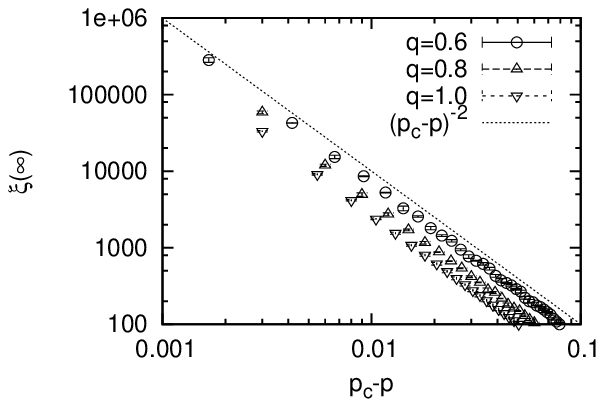} 
\vspace*{0.3cm} \\
\includegraphics[width=6cm]{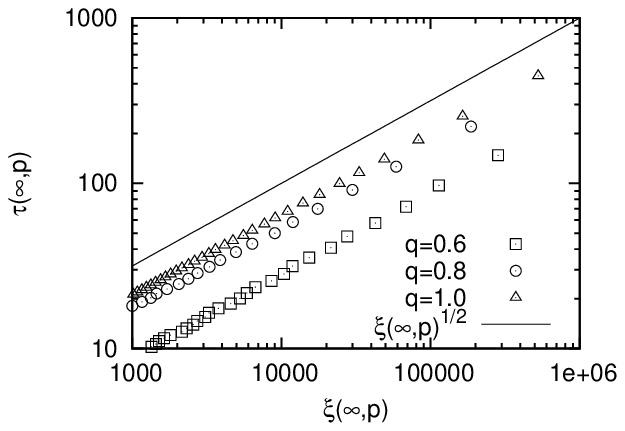} & 
\includegraphics[width=6cm]{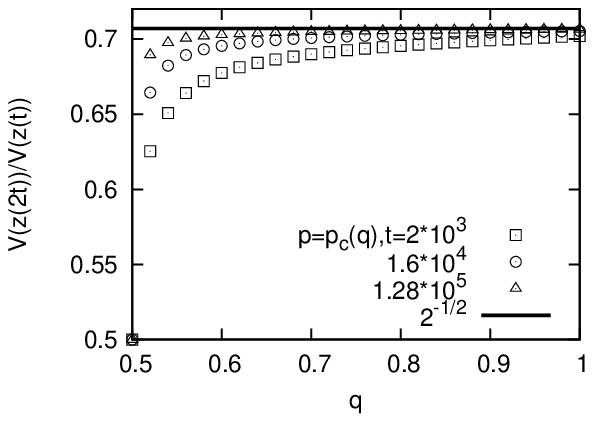} 
\end{tabular}
\end{center}
\caption{\label{fig:Ext_rinf2}
Critical behaviors of $C(t),\xi(\infty),\tau(\infty)$, 
and $\mbox{V}(z(t))$.
Top left: $C(t)/C(0)$ versus $t$ for $p=p_{c}(q)$. 
Top right: $\xi(\infty)$
 versus $p_{c}(q)-p$.
Bottom left: $\tau(\infty)$ versus $\xi(\infty)$ in the critical region
$p\simeq p_{c}(q)$. We adopt $q\in \{0.6,0.8,1.0\}$.  
Bottom right: $\mbox{V}(z(2t))/\mbox{V}(z(t))$ versus 
$q$ at $p=p_{c}(q)$. We adopt 
$t\in \{2\times 10^{3},1.6\times 10^{4},1.28\times 10^{5}\}$.
}
\end{figure}	
We study the critical behaviors of 
the digital model in the limit $r\to \infty$. 
Figure \ref{fig:Ext_rinf2} shows the results.
The top left panel shows $C(t)/C(0)$ versus $t$ for 
$q\in \{0.6,0.8,1.0\}$ and $p=p_{c}(q)$.
One sees that $C(t)$ obeys the power law decay $t^{-1/2}$. 
The top right panel 
shows $\xi(\infty)$ versus $p_{c}-p$ for $q\in \{0.6,0.8,1.0\}$ and 
$p\lesssim p_{c}(q)$.
The critical exponent $\nu_{||}$ for $\xi\propto |p-p_{c}|^{-\nu_{||}}$
 is $\nu_{||}=2$. The lower left panel shows 
$\tau(\infty)$ versus $\xi(\infty)$ for $p\lesssim p_{c}$.
$\nu_{\tau}/\nu_{||}$ is $1/2$.
The lower right panel shows $\mbox{V}(z(2t))/\mbox{V}(z(t))$ versus $q$ 
for $p=p_{c}(q)$. We have no way to extrapolate the results for finite $t$ 
to $\infty$. To see the limit $t\to \infty$, we show the 
results for $t\in \{2\times 10^{3},1.6\times 10^{4},6.4\times 10^{4}\}$.
The plot suggests that the ratio is $1/\sqrt{2}$ in the limit $t\to \infty$
for $q>1/2$.
Because $\gamma_{z}=-\lim_{T\to \infty}\log_{2} \mbox{V}(z(2t))/\mbox{V}(z(t))$,
we have $\gamma_{z}=1/2$.

\section{\label{sec:conclusion}Summaries and Conclusions}
In this paper, we performed a FSS 
analysis of the $r$-th Markov binary
 processes and studied the critical behaviors. 
As the temporal correlation length $\xi$, we propose 
to use the second moment 
correlation time of the autocorrelation function $C(t)$.
As the stochastic process, we consider a mixture of an independent random 
variable and a dependent random variable that depends on the ratio $z$ of 
recent $r$ variables taking 1.
The behavior of the latter dependent variable is defined by $f(z)$.
We consider two types of $f(z)$: (i) analog $f(z)=z$ and (ii) 
digital $f(z)=\theta(z-1/2)$.

\begin{figure}[htbp]
\begin{center}
\includegraphics[width=7cm]{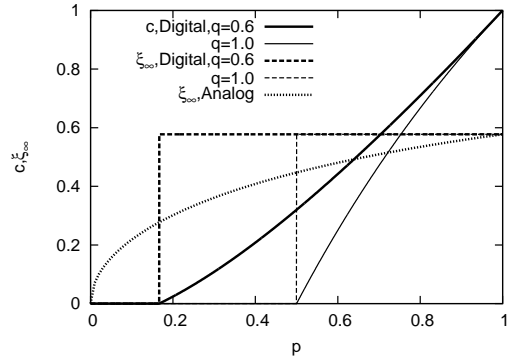}
\end{center}
\caption{\label{fig:Ext_rinf} 
$c$ and $\xi_{\infty}\equiv 
\lim_{t\to\infty}\xi(t)/t$ versus $p$.
}
\end{figure}

We obtained the following results, some of which are 
summarized in Fig. \ref{fig:Ext_rinf} and Table \ref{tab:exponent}. 
% and \ref{tab:summary}.
We denote
the limit value
$\lim_{t\to\infty}\xi(t)/t$ as 
$\xi_{\infty}$.
Figure \ref{fig:Ext_rinf} shows 
$\xi_{\infty}$ and $c$ versus $p$.
Table \ref{tab:exponent} summarizes the critical exponents of the information
cascade phase transition.

\begin{table}[htbp]
\begin{center}
\caption{\label{tab:exponent}
Summary of critical exponents for information 
cascade phase transition.
We denote the analog (digital) model with finite $r$ as 
$A^{r} (D^{r})$ and the limit $r\to \infty$ of $A^{r} (D^{r})$ 
as $A^{\infty} (D^{\infty})$.}
\begin{tabular}{cccccccc}
\hline
Model & $p_{c}$ & $\beta$  & $\alpha$& $\nu_{||}$& $\nu_{\tau}$ & $Z$ & $\gamma_{z}$  \\ 
\hline
$A^{r}$ &1 & $0$  & $0$& $1$& $1$ & $1$ & $0$  \\ 
$D^{r}$ & 1 & $0$  & $0$& $(r+1)/2$& $(r+1)/2$ & $1$ & $0$  \\ 
$A^{\infty}$ & 1 & $0$  & $0$ & NA & NA & $1$ & $0$  \\ 
$D^{\infty},q>1/2$ & 1-1/2q & $1$  & $1/2$& $2$& $1$ & $4/3$ & $1/2$  \\ 
\hline
\end{tabular}
\end{center}
\end{table}

\begin{itemize}
\item Temporal correlation length $\xi$

The critical behavior of the  models obeys the FSS 
 relation based on $\xi$. $\xi$ plays an essential 
 role in the phase transitions and critical behaviors of the systems.

\item Models with $r<\infty$ 

The scaling properties are described by those of the disordered 
system where $C(t)$ decays exponentially \cite{His:2014}. 
The RG stable fixed point is at $\xi_{\infty}=0$, and
the system is in the one-peak disordered phase ($c=0$)
with normal diffusive behavior for $q\neq 1$.
At $p=1$, it is in the  two-peak 
 phase ($c=1$ and $\xi_{\infty}=\sqrt{1/3}$).

\item Analog model in the limit  $r\to \infty$ 
  
The scaling properties are described by those of the 
scale-invariant  
system with $C(t)\propto t^{p-1}$. 
The phase of the system is described by 
$\xi_{\infty}=\sqrt{p/(2+p)}$ and $c=0$.
At $p=1$, it is in the two-peak 
ordered phase ($c=1$ and $\xi_{\infty}=\sqrt{1/3}$).

\item Digital model in the limit  $r\to \infty$ 

 There are two stable and one unstable RG fixed points for 
 $q\neq 1/2$. 
 The two stable fixed points correspond to the 
 one-peak disordered phase ($\xi_{\infty}=0$ and $c=0$)
 and the two-peak phase ($\xi_{\infty}=1/\sqrt{3}$ 
 and $c>0$).
 For $q=1/2$, there is one stable RG fixed point at 
 $\xi_{\infty}=1/\sqrt{3}$, which corresponds to the two-peak phase 
 with $c>0$. $\xi_{\infty}=0$ is an unstable RG fixed point.

\item Finite-size correcting expressions for $\tau/t$ and $\xi/t$ in 
Eq.(\ref{eq:fsc})

\end{itemize}

 We comment on future problems.
 The first problem is the study of a system in the general 
 $f(z)$ case. In our experiment on the information cascade phase transition
 with a two-choice quiz, the behavior of 
 $f(z)$ fell between that of the analog 
 and digital models \cite{Mor:2012}. 
 If we adopt $f(z)=(\tanh(\lambda (z-1/2))+1)/2$ with 
 the control parameter $\lambda$, the system becomes a type of 
 kinetic Ising model in which new spins are added as subjects choose 
 sequentially \cite{His:2014}.
 The important problem is how the scaling relations and critical 
 properties depend on $f(z)$ and the limit $r\to \infty$.
 For a sufficiently slow increase of $r$ with $t$, 
 the system can be equilibrated among recent $r$ variables.
 In the limit $t\to \infty$ with slowly increasing $r$,
 the critical properties might correspond to those of the kinetic Ising model.
 If we take the limit $r\to \infty$ as $r=t$ and $\lambda\to \infty$, 
 the system shows the  phase transition
 for the digital model. 
 Between the two extremes, there might be some rich structure.

 The second problem is the application of the  scaling analysis
 to empirical data.
 In our previous paper, we proposed to use 
 $\gamma_{z}$ to detect the information cascade 
 phase transition \cite{Mor:2012}. 
 Considering that $\xi$ is the essential quantity for 
 characterizing the phase transition and critical properties,
 and $c=\lim_{t\to \infty}C(t)/C(0)$ is the order parameter,
 one should study these quantities. 
 In laboratory experiments, the system size $t$ is severely limited. 
 The finite size
 correcting expressions
 in Eq.(\ref{eq:fsc}) provide 
 convenient expressions for estimating $c$ and 
 $\xi_{\infty}$.

\begin{acknowledgments}
We thank K. Sogo and T. Sasamoto for useful discussions.
This work was supported by Grant-in-Aid for Challenging 
Exploratory Research 25610109. 
\end{acknowledgments}
%\nocite{*}

\bibliographystyle{jpsj} 
\bibliography{myref}% Produces the bibliography via BibTeX.

\providecommand{\noopsort}[1]{}\providecommand{\singleletter}[1]{#1}%
\begin{thebibliography}{10}

\bibitem{Sta:1971}
H.~E. Stanley: {\em Introduction to Phase Transitions and Critical Phenomena}
  (Oxford University Press, London, 1971).

\bibitem{Man:2008}
R.~N. Mantegna and H.~E. Stanley: {\em Introduction to Econophysics:
  Correlations and Complexity in Finance} (Cambridge University Press,
  Cambridge, 2007).

\bibitem{Cas:2009}
C.~Castellano, S.~Fortunato, and V.~Loreto: Rev.Mod.Phys. {\bfseries 81} (2009)
  591.

\bibitem{Gal:2008}
S.~Galam: Int. J. Mod. Phys. C {\bfseries 19} (2008) 409.

\bibitem{Kem:2003}
J.~Kemp: Contemporary Physics {\bfseries 44} (2003) 307.

\bibitem{Boh:2000}
W.~B$\ddot{o}$hm: J. Appl. Prob. {\bfseries 37} (2000) 470.

\bibitem{Pol:1931}
G.P\'{o}lya: Ann. Inst. Henri Poincar\'{e} {\bfseries 1} (1931) 117.

\bibitem{Hod:2004}
S.~Hod and U.~Keshet: Phys. Rev. E {\bfseries 70} (2004) 015104.

\bibitem{His:2006}
M.~Hisakado, K.~Kitsukawa, and S.~Mori: J. Phys. A {\bfseries 39} (2006) 15365.

\bibitem{Usa:2003}
O.V.Usatenko and V.A.Yampol'skii: Phys.Rev.Lett. {\bfseries 90} (2003) 110601.

\bibitem{Val:2007}
P.~Vallois and C.S.Tapiero: Physica A {\bfseries 386} (2007) 303.

\bibitem{Gla:1963}
R.~J. Glauber: J. Math. Phys. {\bfseries 4} (1963) 294.

\bibitem{Pri:1997}
{\em Nonequilibrium Statistical Mechanics in One Dimension}, ed. V.~Privman
  (Cambridge University Press, Cambridge, 1997).

\bibitem{Sta:1999}
H.~E. Stanley: Rev. Mod. Phys. {\bfseries 71} (1999) 358.

\bibitem{New:2005}
M.E.J.Newman: Contem.Phys. {\bfseries 46} (2005) 323.

\bibitem{Hin:2000}
H.~Hinrichsen: Adv.Phys. {\bfseries 49} (2000) 815.

\bibitem{Odo:2004}
G.\'{O}dor: Rev. Mod. Phys. {\bfseries 76} (2004) 663.

\bibitem{Kan:1994}
I.~Kanter and D.A.Kessler: Phys.Rev.Lett. {\bfseries 74} (1995) 4559.

\bibitem{Sta:1995}
H.~Stanley: Nature {\bfseries 378} (1995) 554.

\bibitem{Hui:2008}
T.~Huillet: J.Phys.A {\bfseries 41} (2008) 505005.

\bibitem{Kir:1993}
A.~Kirman: Q. J. Econ. {\bfseries 108} (1993) 137.

\bibitem{Lux:1995}
T.~Lux: Econ. J. {\bfseries 105} (1995) 881.

\bibitem{Kir:2010}
A.~Kirman: {\em Complex Economics: Individual and Collective Rationality}
  (Routledge, 2010).

\bibitem{Alf:2005}
S.Alfarano, T.Lux, and F.Wagner: Comp. Econ. {\bfseries 26} (2005) 19.

\bibitem{Hil:1980}
B.~Hill, D.~Lane, and W.~Sudderth: Ann. Prob. {\bfseries 8} (1980) 214.

\bibitem{Mor:2015}
S.~Mori and M.~Hisakado: arXiv:1501.00764 (2015).

\bibitem{Bik:1992}
S.~Bikhchandani, D.~Hirshleifer, and I.~Welch: J. Polit. Econ. {\bfseries 100}
  (1992) 992.

\bibitem{Dev:1996}
A.~Devenow and I.~Welch: Euro. Econ. Rev. {\bfseries 40} (1996) 603.

\bibitem{Mor:2010}
S.~Mori and M.~Hisakado: J. Phys. Soc. Jpn. {\bfseries 79} (2010) 034001.

\bibitem{And:1997}
L.~R. Anderson and C.~A. Holt: Am. Econ. Rev. {\bfseries 87} (1997) 847.

\bibitem{Kub:2004}
D.~K$\ddot{u}$bler and G.~Weizs$\ddot{a}$cker: Rev. Econ. Stud. {\bfseries 71}
  (2004) 425.

\bibitem{Goe:2007}
J.~Goeree, T.~R. Palfrey, B.~W. Rogers, and R.~D. McKelvey: Rev. Econ. Stud.
  {\bfseries 74} (2007) 733.

\bibitem{Mor:2012}
S.~Mori, M.~Hisakado, and T.~Takahashi: Phys. Rev. E {\bfseries 86} (2012)
  026109.

\bibitem{Mor:2013}
S.~Mori, M.~Hisakado, and T.~Takahashi: J.Phys.Soc.Jpn. {\bfseries 82} (2013)
  0840004.

\bibitem{Lee:1993}
I.~H. Lee: J. Econ. Theory {\bfseries 61} (1993) 395.

\bibitem{His:2010}
M.~Hisakado and S.~Mori: J. Phys. A {\bfseries 43} (2010) 315207.

\bibitem{His:2011}
M.~Hisakado and S.~Mori: J. Phys. A {\bfseries 44} (2011) 275204.

\bibitem{Car:1995}
S.~Caracciolo, R.~G. Edwards, S.~J. Ferreira, A.~Pelissetto, and A.~D. Sokal:
  Phys. Rev. Lett. {\bfseries 74} (1995) 2969.

\bibitem{Bar:1983}
M.~N. Barber. Finite-Size Scaling In C.~Domb and J.L.Lebowitz (eds), {\em Phase Transition and Critical Phenomena}, Vol.~8, pp. 146--268. Academic Press, 1983.

\bibitem{Bin:1985}
K.Binder, M.~Nauenberg, V.~Privman, and A.~P. Young: Phys. Rev. B {\bfseries
  31} (1995) 1498.

\bibitem{Car:1993}
S.~Caracciolo, R.~G. Edwards, A.~Pelissetto, and A.~D. Sokal: Nucl.Phys.B
  {\bfseries 403} (1993) 475.

\bibitem{Sta:1991}
D.~Stauffer and A.~Aharony: {\em Introduction to Percolation Theory}
  (Taylor\&Francis, London, 1991).

\bibitem{His:2014}
M.~Hisakado and S.~Mori: Physica A {\bfseries 417} (2015) 63.

\end{thebibliography}

\appendix
\section{\label{sec:r1}Model for $r=1$}
We derive some quantities of interest for the model for $r=1$.
We write the probabilistic rule for $X(t+1)$ using $X(t)$ as
$\mbox{Prob}(X(t+1)=1)=(1-p)\cdot q+p\cdot X(t)$.
Evaluating the expectation value, we have 
\[
\mbox{E}(X(t+1))=(1-p)q+p\frac{1}{t-1}\cdot {E}(X(t)).
\]
The recursive relation for $\Delta \mbox{E}(X(t))=\mbox{E}(X(t))-q$ is
\[
\Delta \mbox{E(X(t+1))}=p\cdot \Delta \mbox{E}(X(t))
\,\,\, \mbox{with}\,\,\, \Delta \mbox{E}(X(1))=p(1/2-q).
\] 
By solving it, we obtain the result in 
Eq.(\ref{eq:DEXt_r1})

To estimate the correlation function $C(t,t')=\mbox{Cov}(X(t),X(t'))$, 
it is convenient to 
use the transfer matrix. We write the conditional probability 
$\mbox{Pr}(X(s+1)=x_{s}|X(s)=x_{s})$ as the 
$2\times 2$ matrix $T(x_{s+1},x_{s})$.
\[
T(x_{s+1},x_{s})
=\begin{pmatrix}
T(1,1) & T(1,0) \\
T(0,1) & T(0,0) 
\end{pmatrix}
=\begin{pmatrix}
(1-p)q+p & (1-p)q \\
(1-p)(1-q) & (1-p)(1-q)+p 
\end{pmatrix}
\]

Multiplying by $T(x_{s+1}|x_{s})$ for $s=t,\cdots, t'-1,t'>t$, 
we obtain the joint probability function $\mbox{Pr}(X(t')=x_{t'},X(t'-1)=x_{t'-1}|\cdots,X(t)=x_{t})$,
\[
\mbox{Pr}(X(t')=x_{t'},X(t'-1)=x_{t'-1},\cdots |X(t)=x_{t})
=\prod_{s=t}^{t'-1}T(x_{s+1},x_{s}).
\]
Taking the summation over the intermediate variables 
$x_{s},s=t+1,\cdots,t'-1$, we obtain the 
conditional probabilities $\mbox{Pr}(X(t')=x_{t'}|X(t)=x_{t})$ as
\[
\mbox{Pr}(X(t')=x_{t'}|X(t)=x_{t})=T^{|t'-t|}(x_{t'},x_{t})
=\begin{pmatrix}
(1-p^{|t'-t|})q+p^{|t'-t|} & (1-p^{|t'-t|})q \\
(1-p^{|t'-t|})(1-q) & (1-p^{|t'-t|})(1-q)+p^{|t'-t|} 
\end{pmatrix}
\]
We obtain $C(t,t')=\mbox{V}(X(t))\cdot p^{|t'-t|}$ by estimating 
the following equation.
\[
C(t,t')=\mbox{V}(X(t))(T^{|t'-t|}(1,1)-T^{|t'-t|}(1,0)).
\]

$\mbox{E}(z(t))$ is estimated by taking the average of $\mbox{E}(X(s))$ for
$s=1,\cdots,t$ and 
we have 
\[
\mbox{E}(z(t))=q+(\frac{1}{2}-q)p\frac{1-p^{t}}{(1-p)t}
\simeq q+O(t^{-1}).
\]
$\mbox{V}(z(t))$ 
is the summation of $C(s,s'),1\le s,s'\le t$ divided by $t^{2}$.
\begin{eqnarray}
&&\mbox{V}(z(t))
\equiv \frac{1}{t^{2}}\sum_{1\le s,s'\le t}C(s,s')  \nonumber \\
&=&\frac{1}{t^{2}}[t\cdot q(1-q)+(q-1/2)^{2}
(2p\frac{1-p^{t}}{1-p}-p^{2}\frac{1-p^{2t}}{1-p^{2}}) ]
\nonumber \\
&+&\frac{2}{t^{2}}q(1-q)[\frac{p}{1-p}(t-1)-(\frac{p}{1-p})^{2}(1-p^{t-1})]
\nonumber \\
&+&\frac{2}{t^{2}}(q-\frac{1}{2})^{2}\frac{p}{1-p}[
(2p+p^{t+1})\frac{1-p^{t-1}}{1-p}-p^{2}\frac{1-p^{2(t-1)}}{1-p^{2}}-2p^{t}(t-1)] 
\nonumber \\
&\simeq & 
\frac{q(1-q)}{t}\cdot \frac{1+p}{1-p}+O(t^{-2}) 
\nonumber  
\end{eqnarray}
From these results, in the limit $t\to \infty$, $z(t)$ behaves as
\[
z(t) \sim N\left(q,\frac{q(1-q)}{t}
\frac{1+p}{1-p}\right).
\]
To see the relation with the result in \cite{Boh:2000}, 
we change the variable from $z(t)$ to $S(t)$ using $S(t)
\equiv 
\sum_{s=1}^{t}(2X(s)-1)=2t\cdot z(t)-t$. The asymptotic formula for 
$S(t)$ is
\begin{equation}
S(t)\sim N\left((2q-1)t,4t\cdot q(1-q)\frac{1+p}{1-p}\right). \label{eq:Boh}. 
\end{equation}
In \cite{Boh:2000}, the probability for a step in the 
same direction as the previous step is 
denoted as $\alpha$. In our notation, $\alpha=(1-p)1/2+p=(1+p)/2$
 for the symmetric case $q=1/2$. 
After $t$ steps, the probability that 
the random walker stays at $k$  was estimated as
\[
k \sim N(0,t\cdot \alpha/(1-\alpha)).
\]
(To obtain this expression, we set $r=m=t=0$ in 
Theorem 3.1 in \cite{Boh:2000} and change variables from $n$ to $t$.)
 We set $\alpha=(1+p)/2$ and $q=1/2$ in Eq.(\ref{eq:Boh}), and we  
recover the same result. 

\section{\label{sec:A_r}
Analog model with finite $r$}
We write the probabilistic rule for $X(t)$ using $z(t-1,r)$ as
$\mbox{Pr}(X(t)=1)=(1-p)\cdot q+p\cdot z(t-1,r)$.
The expectation value of $X(t)$ obeys the following relation.
\[
\mbox{E}(X(t))=
\begin{cases}
(1-p)q+\frac{p}{t-1}\sum_{s=1}^{t-1}\mbox{E}(X(s)) & t\le r+1, \\
(1-p)q+\frac{p}{r}\sum_{s=t-r}^{t-1}\mbox{E}(X(s)) & t\ge r+2.
\end{cases}
\]
Solving the recursive relation, 
we obtain $\Delta \mbox{E}(X(t))$ for $t\le r+1$ as
\begin{equation}
\Delta \mbox{E}(X(t))=\prod_{s=1}^{t-1}\left(\frac{s-1+p}{s}\right)\Delta \mbox{E}(X(1))\,\,\,,\,\,\,\Delta \mbox{E}(X(1))=p(\frac{1}{2}-q). \label{eq:EX_A}
\end{equation} 
$\mbox{E}(X(t))$ shows a power law
convergence to $q$ with $\Delta \mbox{E}(X(t))\propto t^{p-1}$
if $r$ is large.

For $t\ge r+2$,
%exceeds $r+2$, the effect of
%the variables that are more distant than $r$
%becomes weak, and E$(X(t))$ rapidly converges to $q$.
%For that case
we obtain the following recursive relation.
\[
\Delta \mbox{E}(X(t))-\Delta \mbox{E}(X(t-1))
=\frac{p}{r}(\Delta \mbox{E}(X(t-1))-\Delta \mbox{E}(X(t-r-1))).
\]
We assume the exponential 
convergence of $\Delta E(t)\propto p_{r}^{t}$ and
see that $p_{r}$ obeys the same relation as in Eq.(\ref{eq:pr_A}).

\section{\label{sec:A_rinf}
Analog model in the limit $r\to \infty$}

When we take the limit $r\to \infty$ of the result for the 
analog model with finite $r$, we see that E$(X(t))$ converges to $q$ 
with the power law decay 
$\Delta \mbox{E}(X(t))\propto t^{p-1}$.
Likewise, $C(t)/C(0)$ behaves as 
$C(t)/C(0)\propto t^{p-1}$.

Next, we estimate $C(t,t')=\mbox{Cov}(X(t),X(t'))$ for $t'>t\ge 2$.
We start from the probabilistic rules for $X(t),X(t')$.
$X(t),X(t')$ are coupled through $z(t-1),z(t'-1)$ and 
we have the following relation.
\[
C(t,t')=p^{2}\cdot \mbox{Cov}(z(t-1),z(t'-1))
\]
We decompose $z(t'-1)$ as $z(t'-1)
=\frac{t'-2}{t'-1}z(t'-2)+\frac{1}{t'-1}X(t'-1)$.    
We obtain
\[
C(t,t')=p^{2}\cdot \left(\frac{t'-2}{t'-1}\cdot \mbox{Cov}(z(t-1),z(t'-2))
+\frac{1}{t'-1}\mbox{Cov}(z(t-1),X(t'-1))\right).
\]
Because $\mbox{Prob}(X(t'-1)=1)=(1-p)\cdot q+p\cdot z(t'-2)$, 
the following recursive relation holds.
\begin{equation}
C(t,t')=\frac{t'-2+p}{t'-1}\cdot C(t,t'-1) \label{eq:rec_A}
\end{equation}
For $t'=t+1$, because $\mbox{Cov}(z(t-1),z(t-1))=\mbox{V}(z(t-1))$, we 
have the following relation.
\[
C(t,t+1)=p^{2}\cdot \frac{t-1+p}{t}\cdot \mbox{V}(z(t-1)).
\]
By solving the recursive equation for $C(t,t')$, we obtain
the following expression.
\[
C(t,t')=C(t,t+1)\cdot \prod_{s=t+1}^{t'-1}\frac{s-1+p}{s}.
\]
Because $C(t,t+1) \propto \mbox{V}(z(t))$ and 
$\prod_{s=t+1}^{t'-1}\frac{s-1+p}{s} \sim (t'/t)^{p-1}$, 
we obtain the asymptotic 
behavior of $C(t,t')$ as 
\begin{equation}
\lim_{t,t' \to \infty}C(t,t')
\propto \mbox{V}(z(t))\left(\frac{t'}{t}\right)^{p-1}.
\label{eq:COV_A} 
\end{equation}

\begin{figure}[htbp]
\begin{center}
\includegraphics[width=7cm]{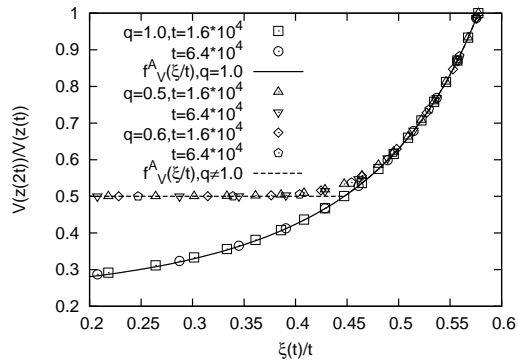}
\end{center}
\caption{\label{fig:rVZ_A_rinf} 
$V(z(2t))/V(z(t))$ versus $\xi(t)/t$ 
for analog model in the limit  $r\to \infty$. 
}
\end{figure}

$\mbox{V}(z(t))$ behaves as in Eq. (\ref{eq:VZ_A}) and  we 
can derive the universal function for $\mbox{V}(z(t))$ as 
\[
\log_{2}f_{V}^{A}(\xi(t)/t)=
\begin{cases}
-1 & \xi(t)/t\le \frac{1}{\sqrt{5}},q\neq 1, \\
\frac{4}{(\xi(t)/t)^{2}-1}-2 & \xi(t)/t>\frac{1}{\sqrt{5}},q\neq 1, \\
\frac{4}{(\xi(t)/t)^{2}-1}-2 & q=1 .
\end{cases}
 \]
At $\xi(t)/t=\frac{1}{\sqrt{5}}$ for $q\neq 1$, there is a 
correction-to-scaling to $f_{V}^{A}$
of order $\log_{t}2$, which is not negligibly small
even for large $t$.

We calculate the ratios $\mbox{V}(z(2t))/\mbox{V}(z(t))$
for the time horizon $t\in \{1.6\times 10^{4},6.4\times 10^{3}\}$ and 
$q\in \{0.5,0.6,1.0\}$.
We plot the ratios as functions of  $\xi(t)/t$ in Fig.\ref{fig:rVZ_A_rinf}.
Because $\gamma_{z}=-\lim_{T\to \infty}\log_{2} \mbox{V}(z(2t))/\mbox{V}(z(t))$ and 
$\xi(t)/t$ is scale invariant,
the figure shows the relation between $\gamma_{z}$ and 
$\lim_{t\to \infty}\xi(t)/t$. 
For $q\neq 1$, there are two phases: the normal diffusion 
phase ($\gamma_{z}=1$) and  
the superdiffusion phase ($\gamma_{z}<1$). 
At $\xi(t)/t=\frac{1}{\sqrt{5}}$, one sees the 
correction-to-scaling to $f_{V}^{A}$, as we have noted previously.
If $q=1$, $\gamma_{z}$ changes smoothly from 2 to 0 as $\xi(t)/t$ 
increases, and the normal-to-superdiffusion 
phase transition does not occur.

%\section{\label{sec:D_rinf}$C(t)$ 
%for digital model in the limit $r\to \infty$ and $q=1$}

\section{\label{sec:R1R2}$c(q,p)$ and $\beta$ for digital model 
in the limit $r\to\infty$}
We have defined $c$ as the difference between the 
conditional probabilities in Eq. (\ref{eq:c_diff}).
We denote the probability that a path starts from the 
wall $n=m+1$ in I, crosses the diagonal only once, and 
reaches the wall $n=m-1$ in II as $R_{1}$.
$R_{2}$ is the probability that the path starts from the wall $n=m-1$
 in II, crosses the diagonal only once, and reaches the wall $n=m+1$ 
in I. The explicit expressions for $R_{1}$ and $R_{2}$ are
\begin{eqnarray}
R_{1}&=&\frac{(1-B)(1\sqrt{1-4A(1-A)}}{A(2-\frac{B}{A}(1-\sqrt{1-4A(1-A)}))},
\label{eq:R1} \\
R_{2}&=&\frac{B(1\sqrt{1-4C(1-C)}}
{(1-C)(2-\frac{1-B}{1-C}(1-\sqrt{1-4C(1-C)}))}. \label{eq:R2}
\end{eqnarray}
Here, $A,B,C$ are defined in Eq.(\ref{eq:p_D}) \cite{His:2011}.
With  $R_{1}$ and $R_{2}$, we can write 
$c(q,p)$ as
\begin{equation}
c(q,p)=p\cdot \frac{(1-R_{1})(1-R_{2})}{1-R_{1}R_{2}} \label{eq:c_exact}.
\end{equation} 
For $p\le p_{c}(q)$, $R_{1}=1$, and we have $c=0$.
If $p>p_{c}(q)$, $R_{1}<1$, and $c$ takes a positive value. 
On the critical line $(p_{c}(q),q)$, $R_{1}=1$, and $R_{2}$ changes smoothly from
1 at $q=1/2$ to $0$ at $q=1$. 
At $p=1$, we have $c=1$. 

We estimate $\beta$ for the infinitesimal perturbation
$(p_{c}(q),q)\to (p_{c}(q)+x,q)$ with $|x|<<1$ as 
$c \propto x^{\beta}$.
The expansion of $R_{1}$ to first order in $x$ is 
\[
R_{1}=1-\frac{16q^{2}x}{4q-1}.
\]
For $q\neq 1/2$, $R_{2}<1$ and $p_{c}(q)>0$. In the expansion of 
$c=p(1-R_{1})(1-R_{2})/(1-R_{1}R_{2})$, 
$(1-R_{1})$ determines the critical exponents.
We obtain $\beta=1$.

\section{\label{sec:Num}Numerical procedure}
We explain the procedure for numerical integration of the master equations.
First, we explain the case of $r\to \infty$.
We denote the joint probability function for $\sum_{s=1}^{t}X(s)$ and $X(1)$ as
$P(t,n,x_{1})\equiv \mbox{Pr}(\sum_{s=1}^{t}X(s)=n,X(1)=x_{1})$.
For $t=1$, $P(1,1,1)=(1-p)q+p/2$ and $P(1,0,0)=(1-p)(1-q)+p/2$. 
The other components are zero.
The master equation for $P(t,n,x_{1})$ is
\begin{equation}
P(t+1,n,x_{1})=q((n-1)/t)\cdot P(t,n,x_{1})+(1-q(n/t))\cdot P(t,n,x_{1})
\end{equation}
Here, we define $q(z)\equiv (1-p)q+p\cdot f(z)$.
We impose the boundary conditions 
$P(t,n,x_{1})=0$ for $n<0$ or $n>t$.
The unconditional probability function $P(t,n)\equiv 
\mbox{Pr}(\sum_{s=1}^{t}X(s)=n)$ is estimated as 
$P(t,n)=\sum_{x_{1}}P(t,n,x_{1})$.

The joint probability function 
$P(x_{t+1},x_{1})=\mbox{Pr}(X(t+1)=x_{t+1},X_{1}=x_{1})$
is then estimated as
\[
P(x_{t+1},x_{1})=
\sum_{n=0}^{t}P(t,n,x_{1})\cdot (q(n/t)\cdot x_{t+1}+(1-q(n/t)\cdot (1-x_{t+1})).
\]
Using $P(x_{t+1},x_{1})$, we can estimate $C(t)$ as
\begin{eqnarray}
C(t)&=&\mbox{E}(X(t+1)X(1))-\mbox{E}(X(1))\cdot \mbox{E}(X(t+1)) \nonumber \\
&=& P(1,1)-\sum_{x_{t+1}}P(x_{t+1},1)\cdot \sum_{x_{1}}P(1,x_{1}). \nonumber
\end{eqnarray}

As for the case $r<\infty$, the procedure is slightly complicated.
First, we define the joint probability functions,  
\begin{eqnarray}
&&P(t,n,x_{t},\cdots,x_{1})\equiv 
\mbox{Pr}(\sum_{s=1}^{t}X(s)=n,X(t)=x_{t},\cdots,X(1)=x_{1}) \,\,\, \mbox{for}\,\,\,t\le r+1 
\nonumber \\ 
&&P(t,n,x_{t},\cdots,x_{t-r+1},x_{1})\equiv 
\mbox{Pr}(\sum_{s=1}^{t}X(s)=n,X(t)=x_{1},\cdots,X(1)=x_{1}) \,\,\, \mbox{for}\,\,\,t\ge r+2. 
\nonumber 
\end{eqnarray}
For $t\le r+1$, $P(t,n,x_{t},\cdots,x_{1})$ depends on all $t$ variables $\{x_{s}\},s=1,2,\cdots,t$.
For $t\ge r+2$, $P(t,n,x_{t},\cdots,x_{t-r+1},x_{1})$ 
depends on the $r$ variables $\{x_{s}\},s=t-r+1,2,\cdots,t$ and $x_{1}$.
Their master equations are
\begin{eqnarray}
P(t+1,n,x_{t+1},\cdots,x_{1})&=&P(t,n,x_{t},\cdots,x_{1})(1-q(n/t))(1-x_{t+1}) \nonumber \\ 
&+& P(t,n-1,x_{t},\cdots,x_{1})q(n/t)x_{t+1} \,\,\, \mbox{for}\,\,\,t\le r.
\nonumber \\
P(t+1,n,x_{t+1},\cdots,x_{t-r+2},x_{1})&=&\sum_{x_{t-r+1}}\left[
P(t,n,x_{t},\cdots,x_{t-r+1},x_{1})(1-q(n/t))(1-x_{t+1})\right. \nonumber \\
&+&P(t,n-1,x_{t},\cdots,x_{t-r+1},x_{1})q(n/t)x_{t+1} \left.\right]
 \,\,\, \mbox{for}\,\,\,t\ge r+1. \nonumber
\end{eqnarray}
We estimate $P(t,n,x_{1})$ from the joint probability functions
 by summing over the variables except $n$ and $x_{1}$.
We then estimate $P(x_{t+1},x_{1})$ and $C(t)$ 
from $P(t,n,x_{1})$.

\end{document}